\documentclass[fleqn,usenatbib]{mnras}

\usepackage{newtxtext,newtxmath}
\usepackage[T1]{fontenc}

\DeclareRobustCommand{\VAN}[3]{#2}
\let\VANthebibliography\thebibliography
\def\thebibliography{\DeclareRobustCommand{\VAN}[3]{##3}\VANthebibliography}

\usepackage{graphicx}	% Including figure files
\usepackage{amsmath}	% Advanced maths commands
\usepackage[dvipsnames]{xcolor}

\title[Formation of polar circumstellar discs]{Formation of  polar circumstellar discs in binary star systems}

\author[Smallwood et al.]{
Jeremy L. Smallwood,$^{1,2}$\thanks{E-mail: jlsmallwood@asiaa.sinica.edu.tw}
Rebecca G. Martin$^2$
and Stephen H. Lubow$^3$
\\
$^1$Institute of Astronomy and Astrophysics, Academia Sinica, Taipei 10617, Taiwan\\
$^2$Department of Physics and Astronomy, University of Nevada, Las Vegas, 4505 South Maryland Parkway, Las Vegas, NV 89154, USA\\
$^3$Space Telescope Science Institute, Baltimore, MD 21218, USA
}

\date{Accepted XXX. Received YYY; in original form ZZZ}

\pubyear{2023}

\begin{document}
\label{firstpage}
\pagerange{\pageref{firstpage}--\pageref{lastpage}}
\maketitle

\begin{abstract}
We investigate the flow of material from highly misaligned and polar circumbinary discs that feed the formation of circumstellar discs around each binary component.  With  three-dimensional hydrodynamic simulations we consider equal mass binaries with low eccentricity. We also simulate inclined test particles and highly-misaligned circumstellar discs around one binary component for comparison. During Kozai-Lidov (KL) cycles, the circumstellar disc structure is altered through exchanges of disc eccentricity with disc tilt.  Highly inclined circumstellar discs and test particles around individual binary components can experience very strong KL oscillations. The continuous accretion of highly misaligned material from the circumbinary disc allows  the  KL  oscillations  of  circumstellar discs to be long-lived. In this process, the circumbinary material is continuously delivered with a high inclination to the lower inclination circumstellar discs. We find that the simulation resolution is important for modeling the longevity of the KL oscillations. An initially polar circumbinary disc forms  nearly polar, circumstellar discs that undergo KL cycles. The gas steams accreting onto the polar circumstellar discs vary in tilt during each binary orbital period, which determines how much material is accreted onto the discs. The long-lived KL cycles in polar circumstellar discs may lead to the formation of polar {\it S}-type planets in binary star systems.
\end{abstract}

\begin{keywords}
binaries: general -- circumstellar matter-- accretion, accretion discs
\end{keywords}

%%%%%%%%%%%%%%%%%%%%%%%%%%%%%%%%%%%%%%%%%%%%%%%%%%

%%%%%%%%%%%%%%%%% BODY OF PAPER %%%%%%%%%%%%%%%%%%

\section{Introduction}
\label{sec::intro}
The majority of stars born in dense stellar clusters are part of binary star systems \citep{Duquennoy1991,Ghez1993,Duchene2013}. The observed orbital eccentricities of binaries vary with orbital separation \citep{Raghavan2010,Tokovinin2016}.  For tight binaries, the eccentricities are small, which implies that there has been circularization of the binary orbit caused by stellar tidal dissipation \citep{Zahn1977}. More widely-separated binaries have observed eccentricities ranging from $e_{\rm b} = 0.39$ to $0.59$, with a considerable number of highly eccentric systems with $e_{\rm b} > 0.8$. The interactions of the binary with surrounding gas may be responsible for the present-day observed binary eccentricities  \citep{Goldreich1980,Artymowicz1991,Artymowicz1992,Armitage2005,Cuadraetal2009,Roedig2011,Munoz2019,Zrake2021}. Circumbinary discs of gas and dust are sometimes observed to be responsible to be providing accreting material onto the binary \cite[e.g.,][]{Alves2019}. The gas flow dynamics from the circumbinary disc onto the binary components has significant implications for planet formation scenarios in binary systems.

Circumbinary discs are commonly observed to be moderately to highly misaligned to the binary orbital plane. For example, the pre-main sequence binary KH 15D has a circumbinary disc inclined by $5-16^\circ$ \citep{Chiang2004,Smallwood2019,Poon2021}. The radial extent of the disc is narrow and presumed to be rigidly precessing to explain the unique periodic light curve. A $\sim 60^\circ$ inclined circumbinary disc is found around the main-sequence binary IRS 43 \citep{Brinch2016}, along with misaligned circumstellar discs around each binary component. There is an observed misalignment of about $70^\circ$ between the circumbinary disc and the circumprimary disc in HD 142527 \citep{Marino2015,Owen2017}. Another young binary, HD 98800 BaBb, has the only observed polar (inclined by $\sim 90^\circ$) gaseous circumbinary disc \citep{Kennedy2019}. The $6$--$10\, \rm Gyr$ old binary system, 99 Herculis, has a nearly polar (about $87^\circ$) debris ring \citep{Kennedy2012,Smallwood2020a}. Apart from binaries, stars may also form in higher-order systems \citep{Tokovinin2014a,Tokovinin2014b}. The circumtriple disc around the hierarchical triple star system, GW Ori, is tilted by about $38^\circ$ \citep{Bi2020,Kraus2020,Smallwood2021GWOri}.
%There are observed differences in the tilts of the circumprimary, circumsecondary, and circumbinary discs. %The system IRS 43 has a configuration where all three discs are misaligned with respect to one another \citep{Brinch2016}. L1551 IRS 5 also has a similar configuration where the circumbinary ring is misaligned by $60^\circ$, and the two circumstellar discs are tilt by at least $45^\circ$ \citep{Fernando2019}. 
%Recently, misaligned circumstellar discs were observed in the young binary system, XZ Tau \citep{Ichikawa2021}. 

%The formation of binary star systems is ubiquitous in the Galaxy \cite[e.g.,][]{Duquennoy1991,Raghavan2010}. 
The observations of inclined circumbinary discs have implications on  planet formation models. Observations from space and ground-based telescopes reveal that $\sim 50$ per cent of the confirmed exoplanets reside in binary systems \citep{Horchetal2014, Deacon2016,Ziegler2018}. For example, the binary system $\gamma$ Cep AB hosts a giant planet around the primary star, $\gamma$ Cep Ab \citep{Hatzes2003}. It is crucial to study the structure and evolution of protoplanetary discs since these are the sites for planet formation \citep{DAngelo2018}.   A forming planet's orbital properties are directly related to the orientation of the protoplanetary disc. For example, the observed young binary system XZ Tau shows both the circumprimary and circumsecondary discs are misaligned to the binary orbital plane \citep{Ichikawa2021}. The binary system HD 142527 shows the presence of a misaligned inner disc around one of the stellar components, presumably fed from the circumbinary disc \citep{Price2018a}. Furthermore, IRAS 04158+2805 is a binary system where the two circumstellar discs and the circumbinary discs have been observed to be misaligned \citep{Ragusa2021}. Therefore, highly-inclined circumstellar discs may give birth to planets on highly-tilted orbits.

Due to viscous dissipation, a misaligned circumbinary disc undergoes nodal precession and evolves towards either a coplanar or polar alignment. For an initially low-inclination circumbinary disc, the disc precesses about the angular momentum vector of the binary and eventually evolves to be coplanar to the binary orbital plane \citep{Facchinietal2013,Foucart2014}.  Slightly misaligned  discs around an eccentric binary undergo tilt oscillations as they align,  due to the nonaxisymmetric potential produced by the eccentric binary \citep{Smallwood2019,Smallwood2020a}.
For highly inclined discs around eccentric orbit binaries, the angular momentum vector of the disc precesses about the eccentricity vector of the binary  \citep[e.g.][]{Aly2015}, which leads the disc to align perpendicular (i.e., polar) to the binary orbital plane \citep{Martinlubow2017,Lubow2018,Zanazzi2018,Martin2018,Cuello2019}. A massive circumbinary disc that is undergoing polar alignment aligns to a generalized polar state which is less than $90^\circ$ \citep{Zanazzi2018,MartinLubow2019,Chen2019}.

Circumbinary gas discs contain a central cavity around the binary where little material is present. The cavity size is determined by where the tidal torque  is balanced with the viscous torque  \citep{Artymowicz1994,Lubow2015,Miranda2015,Franchini2019b,Hirsh2020,Ragusa2020}. The strength of the binary torque on the disc is dependent on the tilt of the circumbinary disc and binary eccentricity.  The tidal torque at a given radius is zero when the circumbinary disc is polar and the binary eccentricity approaches $e_{\rm b} = 1$ \citep{Lubow2018} or if the disc is retrograde \cite[e.g.,][]{Nixon2013}.  In the simplest models, the production of an outward forcing torque by the binary can prevent circumbinary material from flowing through the cavity \citep{LP1974, Pringle1991}. However,  material from the circumbinary disc flows through the binary cavity in the form of gaseous streams \citep[e.g.][]{Artymowicz1996,Gunther2002,NixonKing2012,Shi2012,DOrazio2013,Farris2014,Munoz2019,Alves2019}. These streams are responsible for forming and replenishing circumstellar discs around each binary component. The accretion of material onto the circumstellar discs may aid in the formation of $S$--type planets, those that orbit one component of a binary. Accretion of material onto the central binary may be suppressed for small disc aspect ratios. %, which leads the viscous torque becoming too weak to dominate the tidal torque \cite[e.g.,][]{Ragusa2016, Heath2020}. However, for typical aspect ratios  of protostellar discs \citep[e.g.][]{DAlessio1998}, viscous protoplanetary circcumbinary discs undergo efficient accretion onto the binary.

The structure of a circumstellar disc around one star is strongly affected by the tidal field of the binary companion \citep{Papaloizou1977,Artymowicz1994,Pichardo2005,JangCondell2015}. Circumstellar discs around each binary component undergo tidal truncation. A circumstellar disc in a circular orbit binary is  typically truncated to about one-third to one-half of the binary orbital separation  The tidal truncation radius is expected to decrease with increasing binary eccentricity. 
%%eccentricity of $0.5$  is expected to truncate circumstellar discs to about $0.1-0.15a$ \cite[e.g.,][]{Artymowicz1994}. %The properties of circumstellar discs around members of unequal mass binary are significantly different in size and orientation than discs around an equal-mass binary. 

 Kozai-Lidov (KL) oscillations \citep{Kozai1962, Lidov1962}   have been studied extensively to analyze several astronomical processes involving bodies that orbit a member of a binary system that begin on highly misaligned orbits.
During KL oscillations, the object's inclination is exchanged for eccentricity, and vice versa.
These processes include asteroids and irregular satellites \citep{Kozai1962,Nesvorny2003}, artificial satellites \citep{Lidov1962}, tidal disruption events \citep{Chen2011}, formation of Type Ia supernovae \citep{Kushnir2013}, triple star systems \citep{Eggleton2001,Fabrycky2007}, planet formation with inclined stellar companions \citep{Wu2003,Takeda2005}, giant outbursts in Be/X-ray binaries \citep{Martinetal2014,MartinFranchini2019}, inclined planetary companions \citep{Nagasawa2008}, mergers of binaries in galactic nuclei \citep{Blaesetal2002,Antonini2012,Hamers2018,Hoang2018,Fragione2019a,Fragione2019b}, stellar compact objects \citep{Thompson2011}, and blue straggler stars \citep{Perets2009}.

A highly misaligned initially circular disc around one component of a binary undergoes KL cycles in which its inclination is exchanged for eccentricity, and vice versa \citep{Martinetal2014}. Due to disc dissipation by  viscosity and shocks, these oscillations are   typically significantly damped after a few oscillations.  KL oscillations can occur in a fluid disc with a wide variety of disc and binary parameters \citep{Fu2015}. When the disc becomes eccentric, it overflows its Roche lobe and transfers material to the companion star \citep{Franchini2019}. 
Self-gravity of a disc can suppress disc KL oscillations if the disc is close to being gravitationally unstable \citep{Fu2015b}.  KL oscillations in a circumstellar disc may have significant consequences for planet formation since strong shocks in the gas are produced during high eccentricity phases \citep{Fu2017}.

A misaligned circumbinary disc may form misaligned circumstellar discs around the individual binary components \cite[e.g.,][]{Nixon2013,Smallwood2021}. A highly misaligned disc around one component of a binary may be unstable to the Kozai-Lidov (KL) mechanism \citep{Martinetal2014}. 
 \cite{Smallwood2021} simulated the flow of gas originating from an initially misaligned circumbinary disc by $60^{\circ}$. The misaligned gas streams  that flow into the binary cavity result  in formation  of highly tilted circumstellar discs around each binary component. The inclined circumstellar discs in turn undergo KL oscillations. However, the KL oscillations are long-lived, due to the continuous accretion of inclined material from the circumbinary disc. Long-lived KL cycles have important implications for planet formation in binary systems. 
 
 In this work, we extend the previous study \cite{Smallwood2021} and consider more highly inclined circumbinary discs.
 %We analyze the flow of material originating from a misaligned circumbinary disc, which leads to the formation of circumstellar discs around each binary component. 
 %Highly misaligned circumstellar discs in a binary system can potentially undergo the KL mechanism. 
 We first revisit the dynamics of highly inclined test particle orbits around one component of a binary in Section~\ref{sec::kozai_testpart}. In Section~\ref{sec::setup}, we describe the setup for our hydrodynamical simulations. In Section~\ref{sec::results_CPD}, we discuss the results of our circumprimary disc simulations. We simulate a highly inclined circumprimary disc in a binary 
 %with higher resolution 
 to explore the dynamics of the KL cycles. Previous studies have only dealt with circumprimary disc inclinations $\lesssim 60^\circ$, while we consider higher tilts, including a polar circumprimary disc. In Section~\ref{sec::results_CBD}, we show the results of our hydrodynamical simulations with an initial circumbinary disc, where we consider the flow of material from discs with various initial misalignments, including a polar circumbinary disc.   Finally, a summary is given in Section~\ref{sec::summary}.

\begin{figure} 
\includegraphics[width=\columnwidth]{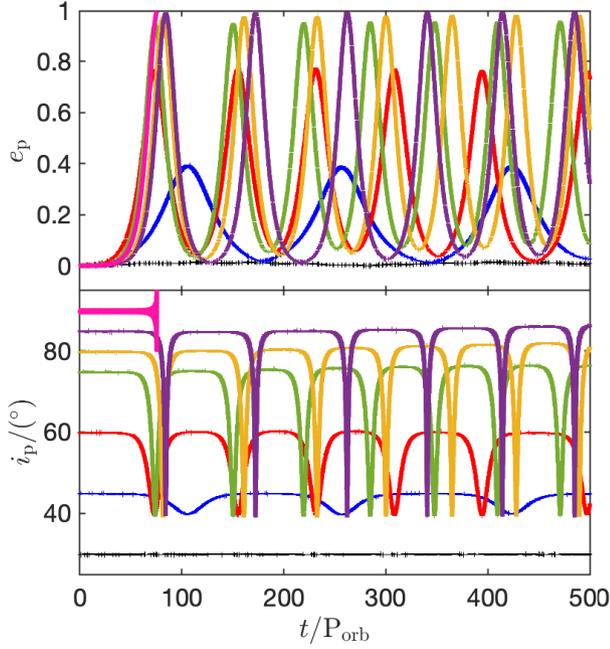}
\centering
\caption{Eccentricity (upper panel) and inclination (lower panel) evolution of circumprimary test particles under the influence of a circular binary  for initially circular orbit particles. We vary the initial particle orbital tilt, $i_0$, beginning with $30^\circ$ (black), $45^\circ$ (blue), $60^\circ$ (red), $75^\circ$ (green), $80^\circ$ (yellow), $85^\circ$ (purple), and $90^\circ$ (pink). The initial orbital radius of the particle is set at $r_0 = 0.06a$, where $a$ is the separation of the binary. The time is in units of binary orbital period $P_{\rm orb}$.}
\label{fig::kozai_particle_i}
\end{figure}

\section{Kozai-Lidov oscillations of test particles}
\label{sec::kozai_testpart}

%As described in the introduction, disc-binary misalignments are commonly observed. If  circumstellar discs have an inclination to the binary orbital plane that is greater than a critical value, the discs undergo the effects of KL oscillations \citep{Martinetal2014}. %These oscillations were first studied with test particles that reside in orbits that are highly misaligned to the orbital plane of a circular binary \citep{Kozai1962,Lidov1962}.
%During these oscillations, the particle exchanges inclination for eccentricity and vice versa.
Before considering discs, we consider the properties of test particle orbits that undergo KL oscillations.
As a consequence of the conservation of the component of the angular momentum that is perpendicular to the binary orbital plane,
 the test particle's inclination is recurrently exchanged for eccentricity. This conservation is expressed as 
\begin{equation}
    \sqrt{1-e^2_{\rm p}}\cos{i_{\rm p}} \approx \rm const,
    \label{eq::ang_mom}
\end{equation}
where $i_{\rm p}$ is the particle inclination with respect to the binary orbital plane and $e_{\rm p}$ is the eccentricity of the test particle. A initially circular orbit particle initially gains eccentricity while reducing its orbital tilt (i.e. going towards alignment which means higher values of $|\cos{i_{\rm p}}|$) and then circularizes while gaining orbital tilt back to its original inclination. For an initially circular orbit particle, KL oscillations only occur if the initial tilt of the test particle $i_{\rm p0}$ satisfies $\cos^2{i_{\rm p0}} < \cos^2{i_{\rm cr}} = 3/5$ \citep{Innanen1997}, which requires that $39^\circ \lesssim i_{\rm p0} \lesssim 141^\circ$. From Eq.~(\ref{eq::ang_mom}), an initially circular particle orbit can achieve a maximum eccentricity given by
\begin{equation}
    e_{\rm max} = \sqrt{1 - \frac{5}{3}\cos^2{i_{\rm p0}}}.
\end{equation}
The increase in a circular particle's eccentricity can be quite significant. For example, if the particle's initial orbit is tilted by $60^\circ$, the maximum eccentricity reached during a KL cycle is about $0.75$.

For eccentric binaries, stronger effects from KL oscillations have been found to exist \citep{Ford2000,Lithwick2011,Naoz2011,Naoz2013a,Naoz013b,Teyssandier2013,Li2014,Liu2015}. The KL oscillation period for a particle in the potential of an eccentric binary is approximately given by
\begin{equation}
    \frac{\tau_{\rm KL}}{P_{\rm b}} \approx \frac{M_{\rm 1} + M_{\rm 2}}{M_{\rm 2}} \frac{P_{\rm b}}{P} (1 - e_{\rm b}^2)^{3/2}
    \label{eq::KL_period}
\end{equation}
\citep{Holman1997,Innanen1997,Kiseleva1998}, where $M_1$ and $M_2$ are the masses of the primary and secondary components of the binary, respectively, $P = 2 \pi/ \sqrt{GM_1/a_{\rm p}^3}$ is the orbital period of the particle with semimajor axis $a_{\rm p}$, $P_{\rm b} = 2\pi / \Omega_{\rm b}$ is the orbital period of the binary, $e_{\rm b}$ is the binary eccentricity, and $\Omega_{\rm b} = \sqrt{G(M_1 + M_2)/a_{\rm b}^3}$ is the binary orbital frequency for binary semimajor axis $a_{\rm b}$.

To simulate an inclined circumprimary test particle in a binary, we use the $N$--body integrator, {\sc MERCURY} \citep{Chambers1999}. The test particle is orbiting the primary companion with an initial tilt $i_0$ relative to the binary orbital plane. The binary components have equal mass so that $M_1 =M_2 = M/2$, where $M$ is the total mass of the binary. \cite{Fu2015b} ran numerous test particle orbits showing the effects the particle and binary parameters have on the induced KL oscillations. Following their work, we model an eccentric inclined particle around one component of an eccentric binary, more applicable to binary systems.

We first simulate an inclined particle in a circular binary to match previous results. Fig.~\ref{fig::kozai_particle_i} shows the eccentricity and inclination of a circumprimary particle as a function of time that begins on a  circular orbit.  
 The analytic solution for these test particle orbits in the quadrupole approximation is given in \cite{Lubow2021}. 
We consider various initial tilts of the test particle orbit.
The critical inclination that the test particle orbit must have to induce KL cycles is $\sim 39^\circ$. Thus, a particle tilt of $30^\circ$ (black line) does not undergo KL oscillations. As the initial inclination of the particle increases, the KL oscillations become more frequent, and the growth in the eccentricity becomes more prominent (in agreement with Fig.~1 in \cite{Fu2015b}).  The trough in the inclination profile of a test particle becomes narrower with initial inclination.  An initial particle orbit tilt of $90^\circ$ becomes unstable and collides with the primary star during the first KL oscillation because the particles eccentricity exceeds $1.0$.  The eccentricity of the polar particle increases almost up to its maximum eccentricity before the tilt begins to change. 

\begin{figure} 
\includegraphics[width=\columnwidth]{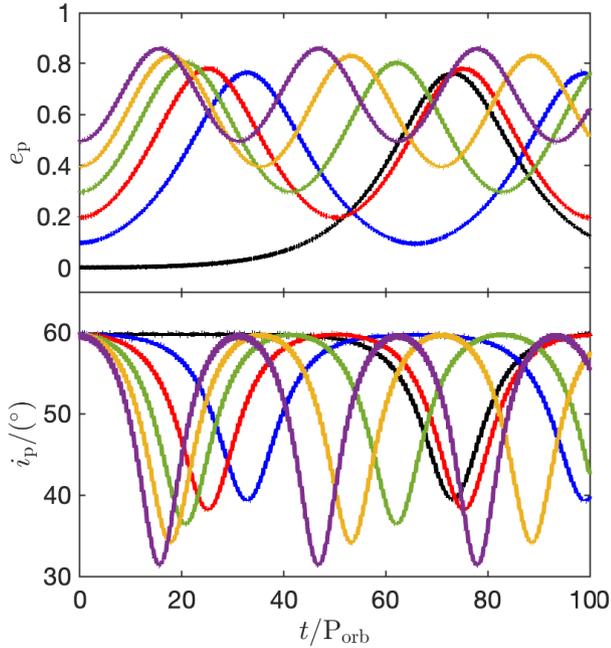}
\centering
\caption{Eccentricity (upper panel) and inclination (lower panel) evolution of circumprimary test particles under the influence of binary with eccentricity $e_{\rm b} = 0.1$. The initial tilt of the particle orbit is set to $60^\circ$.  We vary the initial particle eccentricity $e_0$  beginning with $e_0 = 0$ (black), $0.1$ (blue), $0.2$ (red), $0.3$ (green), $0.4$ (yellow), $0.5$ (purple). The initial orbital radius of the particle is set at $r_0 = 0.06a$, where $a$ is the separation of the binary. The time is in units of binary orbital period $P_{\rm orb}$.} 
\label{fig::kozai_particle_e}
\end{figure}

Next, we set the initial particle tilt to $60^\circ$ around a slightly eccentric binary with  $e_{\rm b} = 0.1$, as we will consider in the disc simulations. We model various initial test particle eccentricities ranging from $0.0$ to $0.5$. Figure~\ref{fig::kozai_particle_e} shows the eccentricity and inclination of eccentric circumprimary particles as a function of time in binary orbital periods. An inclined circular test particle within an eccentric binary has an increased frequency in KL oscillations when compared to a particle  orbiting one component of a circular binary, as expected by equation (\ref{eq::KL_period}).
% Because of the added eccentricity to the binary, the KL oscillation timescale is different from Figure~\ref{fig::kozai_particle_i}.  
From  Figure~\ref{fig::kozai_particle_e}, when the particle eccentricity is increased, the maximum eccentricity reached during a KL oscillation also increases. However, the difference between the initial eccentricity to the maximum eccentricity of the particle decreases as the initial particle eccentricity increases. 

\begin{figure} 
\includegraphics[width=\columnwidth]{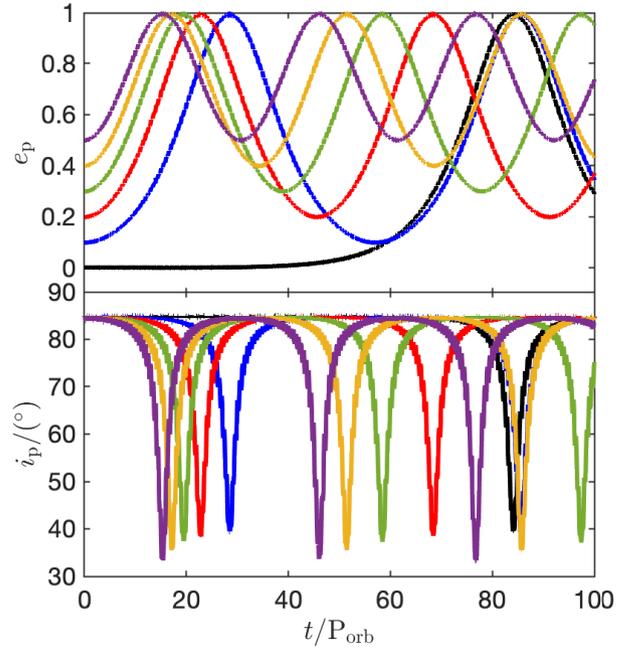}
\centering
\caption{Same as Fig.~\ref{fig::kozai_particle_e} but for nearly polar test particles with an initial orbital tilt $i_0 = 85^\circ$.}
\label{fig::kozai_particle_e2}
\end{figure}

Lastly, we examine the KL mechanism for a nearly polar particle.  From Fig.~\ref{fig::kozai_particle_i}, an initially circular orbit particle with an initial orbital tilt of $85^\circ$ is unstable to KL oscillations but is otherwise stable. %At an initial tilt of $90^\circ$ the test particle orbit becomes non-periodic.
We consider a nearly polar orbit particle with  an initial orbital tilt  $i_0 = 85^\circ$ around a binary with eccentricity $e_{\rm b} = 0.1$.  In Fig.~\ref{fig::kozai_particle_e2} we show the particle eccentricity and inclination as a function of time in binary orbital periods. The various lines correspond to different initial particle eccentricities ranging from $0.0$ to $0.5$. For all values of the initial particle eccentricity we consider, the particle proceeds through KL cycles in a periodic fashion. Unlike the particle beginning at a tilt of $60^\circ$, a nearly polar particle exhibits similar maximum eccentricity  close to unity  during a KL oscillation regardless of initial particle eccentricity. The minimum inclination reached during each KL oscillation is roughly independent of particle initial eccentricity.

\section{Hydrodynamical-simulation setup}
\label{sec::setup}

%We explain the two simulation setup routines for modelling a gaseous circumbinary disc or a circumprimary disc.
We use the smoothed particle hydrodynamics (SPH) code {\sc phantom} \citep{Price2018b} to model gaseous circumbinary and circumstellar discs. {\sc phantom} has been tested extensively for modeling misaligned circumbinary discs \citep{Nixon2012,Nixon2013,Nixon2015,Facchini2018,Smallwood2019,Poblete2019,Smallwood2020a,Aly2020,Hirsh2020,Smallwood2021}, as well as misaligned circumstellar discs around individual binary components \citep[e.g.][]{Martin2014,Dougan2015,Franchini2020}. The suite of simulations is summarised in Table~\ref{table::setup}. In this section we describe the setup  for the binary star, circumprimary disc, and circumbinary disc  in further detail. 

\subsection{Binary star setup}
We model the binary star system as a pair of sink particles, with an initial binary separation $a$. The binary is not static but rather evolves freely in time. Each sink particle is given an initial mass with $M_1$ being the primary mass and $M_2$ being the secondary mass.  The total binary mass is thereby $M = M_1 + M_2$. All of our simulations assume an equal-mass binary ($M_1 = M_2$). In Cartesian coordinates, the orbit of the binary lies in the $x$-$y$ plane initially.  The binary begins initially at apastron along the $x$-axis. The massive sink particles have a hard accretion boundary, meaning that when particles penetrate the sink accretion radius, the particle's mass and angular momentum are deposited onto the star \cite[e.g.,][]{Bate1995}. A large accretion radius is often used to reduce the computation time significantly by neglecting to resolve close-in particle orbits. In this work, however,  we are interested in resolving the formation and evolution of the circumstellar material. Therefore, we adopt a relatively small accretion radius of $0.05a$ for simulations that begin with a circumbinary disc and an accretion radius of $0.025a$ for simulations that begin with a circumprimary disc. Using a smaller accretion radius for the circumprimary disc simulations ensures that the disc lifetime is longer, along with higher disc resolution.
The more eccentric the binary, the smaller the outer truncation radius for the circumstellar discs \citep{Artymowicz1994}. Having a small binary eccentricity helps with the resolution of the circumstellar discs.
On the other hand, to have a stable polar circumbinary disc, the binary eccentricity needs to be a non-zero value. The initial binary eccentricity is set to $e_{\rm b} = 0.1$, with the binary eccentricity vector along the positive $x$--axis.  With this value of binary eccentricity, the critical tilt of the circumbinary disc to remain nearly polar is $\sim 77^\circ$ \cite[see eq. 33 in ][]{MartinLubow2019}.

\begin{table*}
	\centering
	\caption{The setup of the SPH simulations that includes an initial circumprimary disc (CPD) or circumbinary disc (CBD). The table lists the initial parameters beginning with the disc tilt $i_0$,  inner disc radius $r_{\rm in}$,  outer disc radius $r_{\rm out}$, $\alpha$ viscosity parameter, disc aspect ratio at inner disc radius $H/r_{\rm in}$, disc aspect ratio at outer disc radius $H/r_{\rm out}$, the  number of particles, and whether or not the circumstellar discs undergo the Kozai-Lidov (KL) instability. }
%	\label{tab:example_table}
	\begin{tabular}{lccccccccc} % four columns, alignment for each
		\hline
	    Model & Disc Setup  & $i_0/^\circ$ & $r_{\rm in}/a$ & $r_{\rm out}/a$ & $\alpha$ & $H/r_{\rm in}$  & $H/r_{\rm out}$ & $\#$ Particles & KL unstable?\\
		\hline
		\hline
		run1 & CPD  & $60$ & $0.025$ & $0.25$ & $0.01$  & $0.035$ & $0.02$  & $750,000$ & Yes\\
        run2 & CPD & $70$ & $0.025$ & $0.25$ & $0.01$  & $0.035$ & $0.02$  & $750,000$ & Yes \\
        run3 & CPD  & $80$ & $0.025$ & $0.25$ & $0.01$  & $0.035$ & $0.02$  & $750,000$ &  Yes \\
        run4 & CPD & $90$ & $0.025$ & $0.25$ & $0.01$   & $0.035$ & $0.02$  & $750,000$ & Yes \\
        run5 & CPD  & $100$ & $0.025$ & $0.25$ & $0.01$   & $0.035$ & $0.02$  & $750,000$ &  Yes \\
		\hline
% 		run6 & CBD  & $0.5$ & $0.5$ & $0$ & $2$ & $3$ & $0.1$ & $0.09$ & $1.5\times 10^6$ & No \\
% 		run7 & CBD & $0.5$ & $0.5$ & $20$ & $2$ & $3$   & $0.1$  & $0.09$  & $1.5\times 10^6$ & No \\
%         run8 & CBD & $0.5$ & $0.5$ & $40$ & $2$ & $3$  & $0.1$  & $0.09$  & $1.5\times 10^6$ & Yes \\
     	run6$^*$ & CBD  & $60$ & $1.6$  & $2.6$ & $0.1$   & $0.1$  & $0.088$  & $1.5\times 10^6$ & Yes \\
     	run7 & CBD & $60$ & $1.6$  & $2.6$  & $0.1$   & $0.1$  & $0.088$  & $750,000$ &  Yes \\
        run8 & CBD & $90$ & $1.6$ & $2.6$ & $0.1$  &$0.1$  & $0.088$  & $1.5\times 10^6$ & Yes  \\

%        run9 & CBD & $0.2$ & $0.5$ & $0.5$ & $90\degree$ & $1.6$ & $2.6$   &  $0.1$ & $0.088$  & $1.5\times 10^6$ \\
%        run10 & CBD & $0.3$ & $0.5$ & $0.5$ & $90\degree$ & $1.6$ & $2.6$   & $0.1$  & $0.088$  & $1.5\times 10^6$ \\
        % run12 & CBD  & $0.55$ & $0.45$ & $90$ & $1.6$  & $2.6$   & $0.1$  & $0.088$  & $1.5\times 10^6$ & Yes \\
		\hline
        \multicolumn{10}{l}{$^{*}$ \text{Simulation from 
        \citet{Smallwood2021}}}\\
        \hline
	\end{tabular}
    \label{table::setup}
\end{table*}

\subsection{Circumprimary disc setup}
To model a circumprimary disc, we follow the methods of \cite{Martin2014}. Runs 1-5 in Table~\ref{table::setup} simulate initially a circumprimary disc. The inner and outer disc radii are set at $r_{\rm in} = 0.025a$ and $r_{\rm out} = 0.25a$, respectively, with a initial total disc mass $M_{\rm CPD} = 10^{-3} M$.  The circumprimary disc consists of $750,000$ equal-mass Lagrangian particles. We neglect any effects of self-gravity. The disc surface density profile is initially a power law distribution given by
 \begin{equation}
     \Sigma(r) = \Sigma_0 \bigg( \frac{r}{r_{\rm in}} \bigg)^{-p},
     \label{eq::sigma}
 \end{equation}
 where we set $p = 3/2$.  We adopt a locally isothermal disc with sound speed $c_{\rm s} \propto R^{-3/4}$, $H/r = 0.035$ at $r = r_{\rm in}$, and $H/r = 0.02$ at $r = r_{\rm out}$. With this prescription, the viscosity parameter $\alpha$ and $\langle h \rangle / H$ are effectively constant over the radial extend of the disc \citep{Lodato2007}. For the circumprimary disc simulations, we take the \cite{Shakura1973} $\alpha$ parameter to be $0.01$. 
 To accomplish this, the SPH artificial viscosity coefficients are set as $\alpha_{\rm AV} = 0.18$ and $\beta_{\rm AV} = 2.0$. The disc is resolved with shell-averaged smoothing length per scale height $\langle h \rangle / H \approx 0.55$.

\subsection{Circumbinary disc setup}
To model an initially flat but tilted gaseous circumbinary disc, we follow the methods of \cite{Smallwood2021}. Runs 6, 7, and 8 in Table~\ref{table::setup} describe the simulations of a circumbinary disc. The disc initially consists of $1.5\times10^6$ equal-mass Lagrangian SPH particles. We also model a $750,000$ particle simulation for a resolution study. The simulations run for $45\, P_{\rm orb}$, where $P_{\rm orb}$ is the orbital period of the binary. This is sufficient time for the forming circumstellar discs to reach a quasi-steady state. We simulate initially highly misaligned disc inclinations of $i_0 = 60^\circ,90^\circ$.  A disc with $i_0 = 90^\circ$ is in a polar configuration, where the angular momentum vector of the disc is aligned to the eccentricity vector of the binary.  At the beginning of our simulations, we select an initial inner disc radius, $r_{\rm in}$, and outer disc radius, $r_{\rm out}$, where the initial total disc mass, $M_{\rm CBD}$, is confined. All of the simulations model a low-mass circumbinary disc such that $M_{\rm CBD} = 10^{-3}M$. We choose the circumbinary disc to be radially very narrow and close to the binary orbit. This is done to maximise the accretion rate onto the binary  and hence the resolution of the circumstellar discs \cite[e.g.,][]{Smallwood2021}. 
%For simulations with $i_0 \leq 40^\circ$, we take $r_{\rm in} = 2a$ and $r_{\rm out} = 3a$. 
For our simulations, we take $r_{\rm in} = 1.6a$ and $r_{\rm out} = 2.6a$.
The tidal torque is weaker at a given radius for a more highly misaligned disc which allows the inner disc radius to lie closer to the binary than a coplanar disc \cite[e.g.,][]{Lubow2015,Miranda2015,Lubow2018}.  The inner truncation radius of a polar circumbinary disc is around $1.6\,a$ \citep{Franchini2019b}, much smaller than the $2-3\, a$ expected for coplanar discs \citep{Artymowicz1994}.

The disc surface density profile follows from Equation~(\ref{eq::sigma}). The physical disc viscosity is incorporated by using artificial viscosity $\alpha^{\rm av}$, which is detailed in \cite{Lodato2010}.  By using our surface density profile and a disc aspect ratio $H/r = 0.1$ at $r_{\rm in}$, the shell-averaged smoothing length per scale height $\langle h \rangle / H$ and the disc viscosity parameter $\alpha$ are constant over the radial extent of the disc \citep{Lodato2007}. The circumbinary disc is initially resolved with $\langle h \rangle / H \approx 0.11$. 
The parameters for the simulations require a high viscosity in order to {\bf maximize the accretion rate on to the circumstellar discs and} provide better resolution.   We consider a relatively high value for the \cite{Shakura1973} $\alpha_{\rm SS}$ of $0.1$. In a more realistic system, the disc viscosity may be lower.

In order to more accurately simulate the formation and development of circumstellar discs, we adopt the locally isothermal equation of state of \cite{Farris2014} and set the sound speed $c_{\rm s}$ to be
\begin{equation}
    c_{\rm s} =  {\cal{F}} c_{\rm s0}\bigg( \frac{a}{M_1 + M_2}\bigg)^q \bigg( \frac{M_1}{r_1} + \frac{M_2}{r_2}\bigg)^q,
    \label{eq::EOS}
\end{equation}
where $r_1$ and $r_2$ are the radial distances from the primary and secondary stars, respectively, and
 $c_{\rm s0}$ is a constant with dimensions of velocity. 
  $q$ is set to 3/4. ${\cal{F}}$ is a dimensionless  function of position that we define below. This sound speed prescription guarantees that the temperature profiles in the circumprimary and circumsecondary discs are set by the primary and secondary stars, respectively. For $r_1, r_2 \gg a$, $c_{\rm s}$ is set by the distance from the binary centre of mass. 

To increase the resolution of the circumstellar discs, we include a function $\cal{F}$ in Equation (\ref{eq::EOS}) as detailed in \cite{Smallwood2021}. The purpose of $\cal{F}$ is to modify the sound speed around each binary component so that the viscous timescale is longer. This increases the mass (and hence the resolution) in the steady-state circumstellar discs. We take
\begin{equation}
  \cal{F}=\begin{cases}
   \sqrt{0.001}, & \text{if $r_1\, {\rm or}\, r_2 < r_{\rm c}$},\\
    1, & \text{otherwise},\\
  \end{cases}
\end{equation}
where $r_{\rm c}$ is the cutoff radius. We set a cutoff radius of $r_{\rm c}=0.35a$ from each binary component \cite[e.g.,][]{Smallwood2021}. Using the prescription mentioned above ensures that the disc aspect ratio of the circumstellar discs at radius $r=0.1a$ is $H/r \sim 0.01$, which is one-tenth of the disc aspect ratio at the initial inner circumbinary disc radius.

\subsection{Analysis routine}
 We analyse the disc and binary parameters as a function of time.  The parameters include tilt, eccentricity, the longitude of the ascending node, mass, and mass accretion rate.  To probe the circumprimary disc simulations, we average over particles in the radial range from $0.025a$ to a distance of $0.30a$. For the circumbinary disc simulations,  we average over particles in the radial range from $1.4a$ to a distance of  $10a$. For the forming circumstellar discs, we average over all particles bound to each binary component (i.e., the specific energies, kinetic plus potential, of the particles are negative, neglecting the thermal energy).  
 The tilt, $i$, is defined as the angle between the initial angular momentum vector of the binary (the $z$-axis) and the angular momentum vector of the disc. The longitude of the ascending node, $\phi$, is measured relative to the $x$-axis (the initial binary eccentricity vector). 

%\RGM{Are all the figures for the CBDs averaged over the whole disc? If they are perhaps you don't need to talk about the grid?}
%For each circum-single disc, we assess the mean properties of the particles, such as the surface density, inclination, longitude of ascending node, and eccentricity. 

\begin{figure} 
\includegraphics[width=\columnwidth]{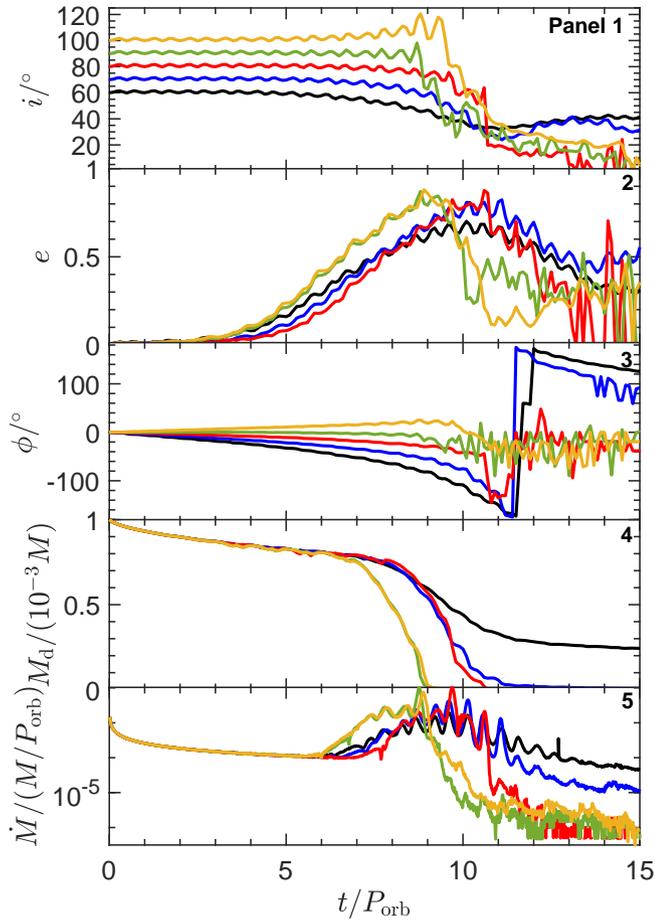}
\centering
\caption{ Evolution of a KL unstable circumprimary disc as a function of time in units of the binary orbital period $P_{\rm orb}$.  We simulate five different initial disc inclinations, which are $60^\circ$ (run1 from Table~\ref{table::setup}, black), $70^\circ$ (run2, blue), $80^\circ$ (run3, red),  $90^\circ$ (run4, green), and $100^\circ$ (run5, yellow).  The disc parameters are tilt $i$ (panel 1), eccentricity $e$ (panel 2), longitude of the ascending node $\phi$ (panel 3), and disc mass $M_{\rm d}$ (panel 4). The mass accretion rate $\dot{M}$  onto the primary star is shown in panel 5. }
\label{fig::kozai_disc}
\end{figure}

\begin{figure} 
\begin{center}
\includegraphics[width=0.49\columnwidth]{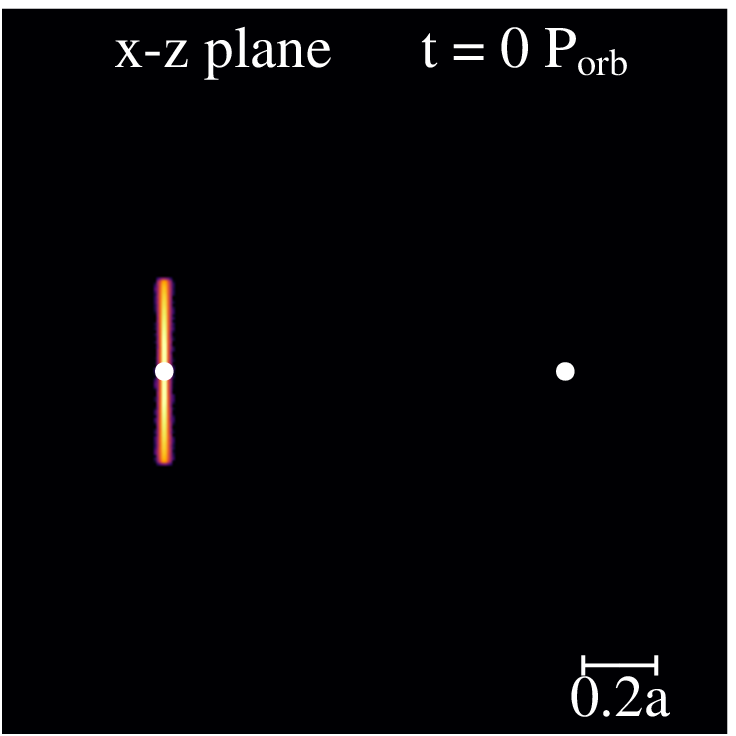}
\includegraphics[width=0.49\columnwidth]{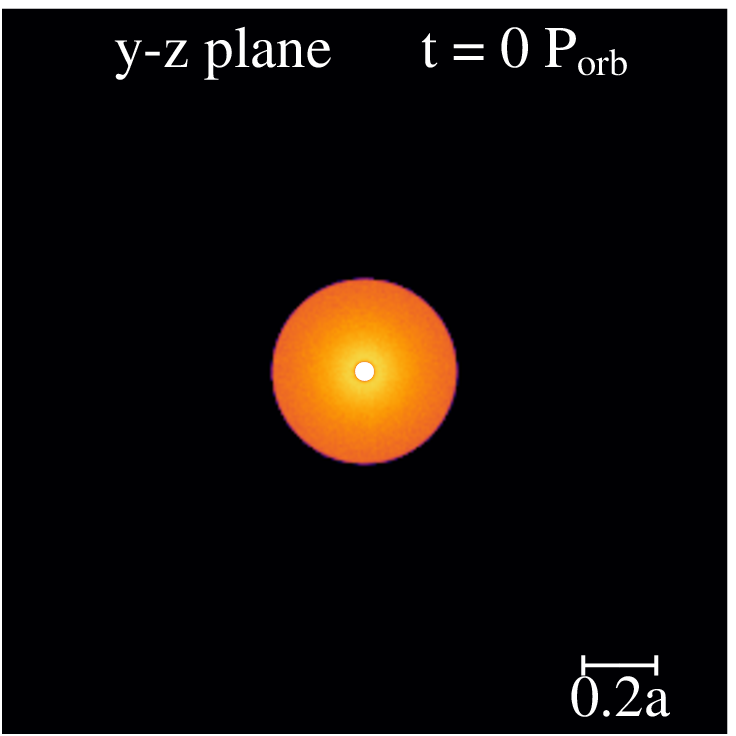} 
\includegraphics[width=0.49\columnwidth]{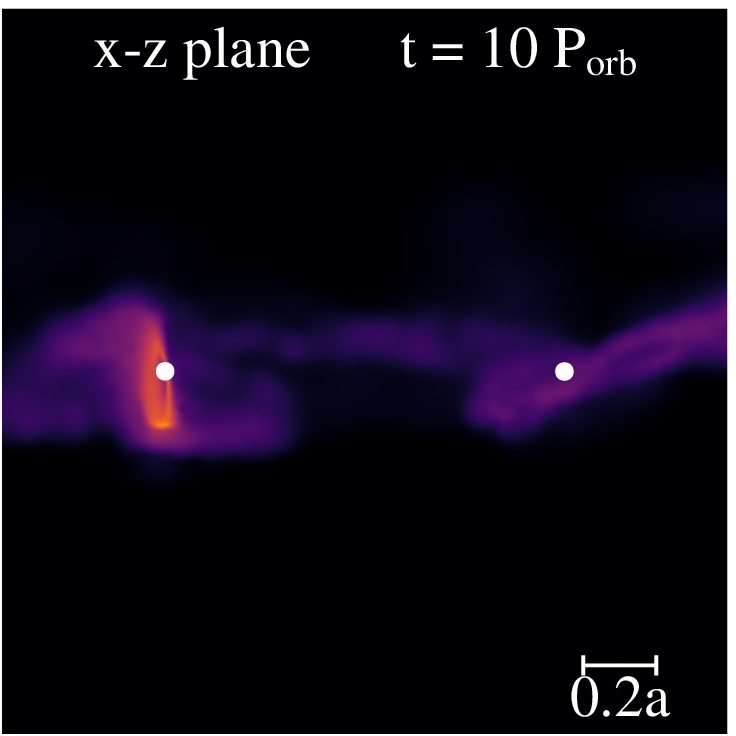} 
\includegraphics[width=0.49\columnwidth]{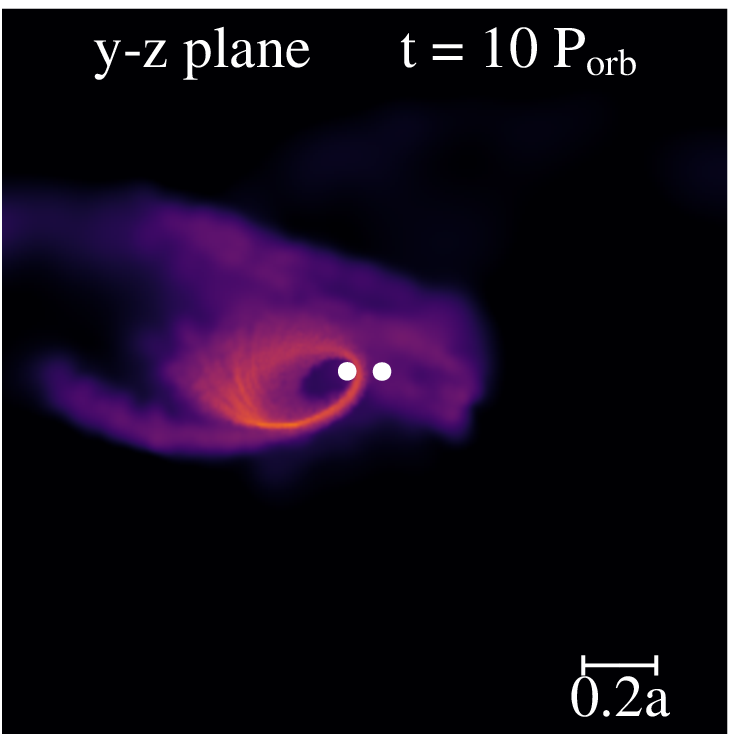} 
\includegraphics[width=0.49\columnwidth]{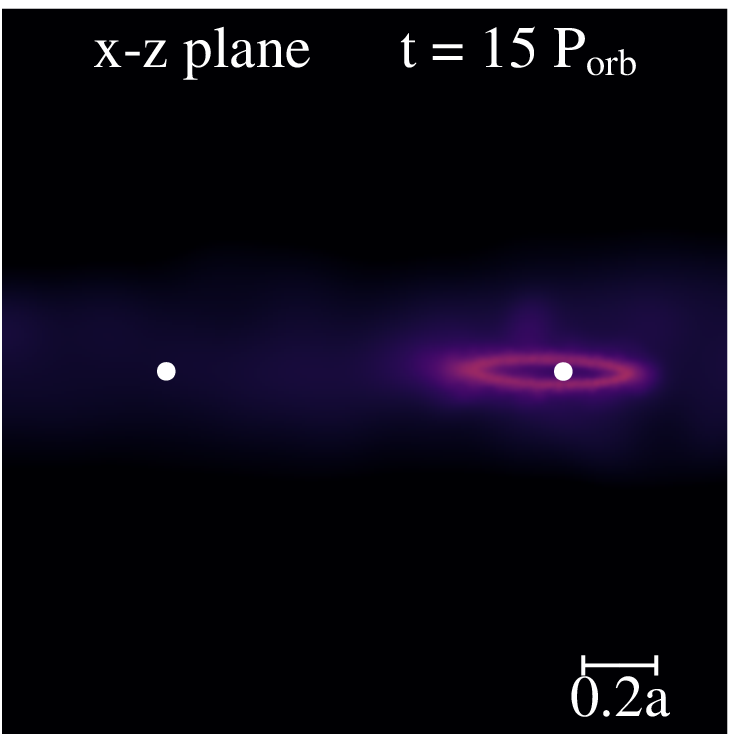} 
\includegraphics[width=0.49\columnwidth]{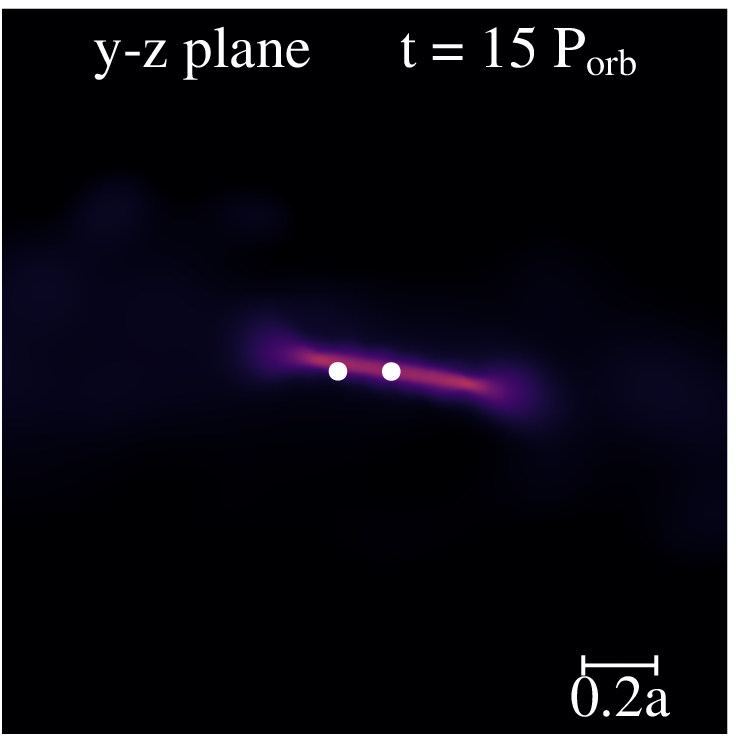} 
\end{center}
\caption{The evolution of polar circumprimary disc (run4 from Table~\ref{table::setup}).  The white circles denote the eccentric orbit binary components with an initial binary separation of $a$. The top row shows the initial disc setup. The middle and bottom rows show the disc evolution at $t = 10\, P_{\rm orb}$ and $t = 15\, P_{\rm orb}$, respectively,  where $P_{\rm orb}$ is the binary orbital period. The color denotes the gas surface density, with the orange regions being about three orders of magnitude larger than the purple regions. The left column shows the $x$--$z$ plane, and the right column shows the $y$--$z$ plane. At $t = 10\, P_{\rm orb}$, the circumprimary disc is highly eccentric due to the Kozai-Lidov instability. Also, at this time, a circumsecondary disc is being formed from material flowing close to the secondary binary component from the eccentric circumprimary disc. At $t = 15\, P_{\rm orb}$, the circumprimary disc has completely dissipated from being accreted onto the primary star and  transferring material to the secondary star. At this time, there is more material in the newly formed circumsecondary disc. }
\label{fig::splash_90pri}
\end{figure}

\begin{figure} 
\includegraphics[width=\columnwidth]{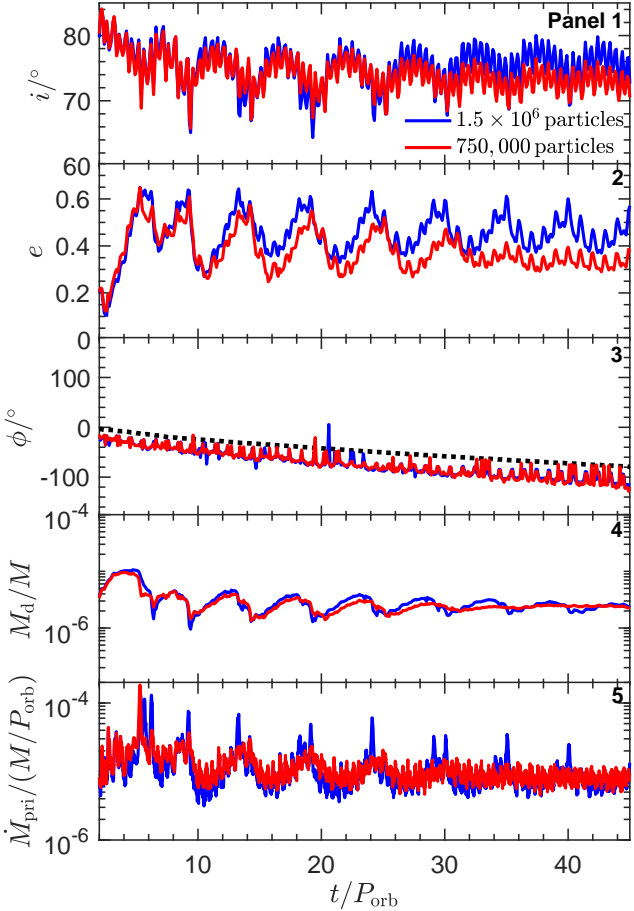}
\centering
\caption{Resolution study for a circumbinary disc that is initially misaligned by $60^\circ$. The blue curves represent the simulation with initially $1.5\times 10^6$ particles in the circumbinary disc, while the red curves denotes the simulation with initially $750,000$ particles. The first four panels show the disc parameters for the newly forming circumprimary disc as a function of time in units of the binary orbital period, $P_{\rm orb}$. The disc parameters are tilt $i$ (panel 1), eccentricity $e$ (panel 2), longitude of the ascending node $\phi$ (panel 3), and  disc mass $M_{\rm d}$ (panel 4). The black dotted curve in the third panel denotes the circumbinary disc. The lower panel shows the mass accretion rate onto the primary star $\dot{M}_{\rm pri}$ (panel 5). }
\label{fig::i60}
\end{figure}

\section{Hydrodynamical results with a circumprimary disc}
\label{sec::results_CPD}
This section  considers the evolution of a circumprimary disc in the absence of accretion from a circumbinary disc. This enables us to  disentangle
the effect of accretion onto the circumstellar discs.  We focus on large circumprimary disc misalignments in an eccentric binary star system. 
We consider five different initial disc tilts, $60^\circ$ (run1 from Table~\ref{table::setup}), $70^\circ$ (run2), $80^\circ$ (run3), $90^\circ$ (run4),  and $100^\circ$ (run5). Figure~\ref{fig::kozai_disc} shows the disc tilt, eccentricity, the longitude of the ascending node, the mass of the circumprimary disc, and the accretion rate onto the primary star as a function of time in binary orbital periods. The disc exhibits KL cycles for each initial tilt, where the disc eccentricity and inclination are exchanged. For a disc with an initial tilt of $60^\circ$, \cite{Martinetal2014} found that the first KL oscillation occurred around $10\, \rm P_{\rm orb}$ for a circular binary. In our case, the disc with the same initial tilt undergoes the first KL oscillation much sooner  due to the binary having a  slightly eccentric orbit \cite[see Fig.12 in][]{Fu2015}. 
Due to viscous dissipation and the lack of circumbinary material, the KL oscillations damp quickly in time. For higher initial inclinations, $70^\circ$, $88^\circ$, $90^\circ$ and $100^\circ$, the discs do not survive after one KL oscillation  for our given sink size. The discs become very eccentric, which leads to the majority of the disc material being accreted by the primary star. Increasing the resolution of these simulations does not lengthen the disc lifetime. However, if we were to use a smaller sink size, then the disc could survive through the KL oscillations.  A smaller sink size would ensure that a larger portion of the disc could survive. An accretion radius of $\sim 0.01\, \rm au$ is comparable to the size of the star, but we simulate a larger sink size for computational reasons and to compare with the circumbinary disc simulations detailed in the next Section.
The initially polar disc's tilt does not change much from polar before the majority of the disc is accreted.  This is likely a consequence of the high disc eccentricities that are developed which is consistent with the results for test particle orbits (see Fig.~\ref{fig::kozai_particle_i}). In the retrograde case, $i_0 = 100^\circ$, as the disc eccentricity increases, the inclination also increases,  opposite to the prograde cases.

 Highly inclined particle orbits experience a  large (nearly $180^\circ$)  shift in $\phi$ within a small time interval centered about the eccentricity maximum (see the plot for $\Omega(t)$
in Figure 1 of \cite{Lubow2021}). This large shift does not appear in Figure~\ref{fig::kozai_disc} or in any of our other phase results.
We are not sure why this is the case. Perhaps the disc is unable to respond to such a large shift
within a short time.

We further examine the evolution of the polar ($i_0 = 90^\circ$) circumprimary disc. In Fig.~\ref{fig::splash_90pri}, we show the polar circumprimary disc structure at three different times, $t=0\, P_{\rm orb}$, $10\, P_{\rm orb}$, and $15\, P_{\rm orb}$.  Initially, the polar disc around the primary star (left white dot) is edge-on in the $x$-$z$ plane and face-on $y$-$z$ plane. At $t=10\, P_{\rm orb}$, the disc is at peak eccentricity growth from the KL instability. Also, at this time, streams of material from the circumprimary disc flow around the secondary star  (right white dot) and begin forming a circumsecondary disc. At $t=15\, P_{\rm orb}$, the circumprimary disc has dissipated due to accretion onto the primary star and transporting material to the circumsecondary disc. The newly formed circumsecondary disc is at a lower tilt,  below the threshold, to induce the KL cycles.

\section{Hydrodynamical results with a circumbinary disc}
\label{sec::results_CBD}

In this section we examine how misaligned and polar circumbinary material flows through the binary cavity and forms circumstellar discs around each binary component. We first conduct a resolution study of our earlier work from \cite{Smallwood2021}, modeling an initially $60^\circ$ misaligned circumbinary disc. We then focus on the polar circumbinary disc case.

\subsection{Resolution Study}
We examine a circumbinary disc with an initial misalignment of $i_0 = 60^\circ$ with two different initial numbers of particles, $1.5\times 10^6$ (run6) and $750,000$ (run7). The upper four panels in Figure~\ref{fig::i60} show the circumprimary disc parameters as a function of time. The bottom panel shows the mass accretion rate onto the primary star. The blue curves represent the  $1.5\times 10^6$ particle simulation, while the red curves represent the $750,000$ particle simulation. Panels 1 and 2 show the evolution of disc eccentricity and inclination where the forming circumprimary disc undergoes KL oscillations from the continuous accretion of material from the circumbinary disc. The oscillations damp in time at both resolutions, with the lower resolution simulation damping more quickly. Therefore, the oscillations are likely limited by resolution. If the accretion timescale is long compared to the KL timescale, we expect the KL oscillations to damp over time,  similar to the circumprimary disc simulations without accretion shown in the previous Section.  If the accretion timescale is short compared to the KL timescale, there should be no KL oscillations present.  In this case, the material moves through the disc faster than it becomes unstable to KL oscillations. We expect the optimal oscillations when the timescales are comparable because the disc refills mass on the timescale that the oscillations take place. For the simulation with a $60^\circ$ tilted circumbinary disc, the accretion timescales for the primary and secondary are $\sim 1.5\, \rm P_{orb}$, whereas the KL timescale for this simulation is $\sim 5\, \rm P_{orb}$. The simulation is in the regime where the accretion timescale is shorter than the KL oscillation timescale because when the disc becomes eccentric during the KL oscillations, a large amount of disc material is accreted, reducing the accretion timescale. 
 However, the accretion timescale is dependent on the disc viscosity. In our hydrodynamical simulations, we use an artificial viscosity to model an expected \cite{Shakura1973} viscosity coefficient. The number of Lagrangian particles determines how close the artificial viscosity is to the actual value. Thus, the $\alpha$ is artificially higher at lower resolutions, leading to a shorter accretion timescale. For our higher-resolution simulation, the $\alpha$ is lower, leading to a longer accretion timescale.

Panel 3 in Fig.~\ref{fig::i60} shows the longitude of the ascending node as a function of time. The precession rate of the circumprimary disc is only slightly faster than the circumbinary disc on average. 
In the absence of the effects of KL oscillations, the nodal precession rate of the primary disc, assuming constant surface density $\Sigma$ out to disc radius $r$ from the primary,
is given by
\begin{equation}
    \omega_{\rm pr} = - \frac{15M_2 r^3}{32 M_1 a_{\rm b}^3}\cos{(i)} \, \Omega(r),
    \label{omnonKL}
\end{equation}
 where $i$ is inclination angle of the primary disc relative to the binary orbital plane and $\Omega = \sqrt{G M_1/r^3}$ is the angular velocity in the disc \citep{Larwoodetal1996}. With $r =  0.35\, a_{\rm b}$, we find $\omega_{\rm pr} = 6^\circ/P_{\rm orb}$ with a revolution period of $\sim 56\, \rm P_{orb}$. Therefore, the circumstellar discs should have nodally precessed $75$ per cent of a revolution in $45\, \rm P_{\rm orb}$. In panel 3 we see that the circumstellar discs have only completed roughly $30$ per cent of a nodal revolution. It is possible that 
the circumprimary phase  is affected by the phase of accreted gas from the circumbinary disc that undergoes relatively slow nodal precession. As discussed in Section  \ref{sec::results_CPD}, KL oscillations modify the nodal precession rate of a test particle in a way that we do not see in the disc simulations.  Lastly, the mass in the circumprimary discs oscillates in time, with the troughs corresponding with each high eccentricity period. 
During each high eccentricity phase, the accretion rate peaks as seen in panel 5.

\begin{figure} 
\begin{center}
\includegraphics[width=0.49\columnwidth]{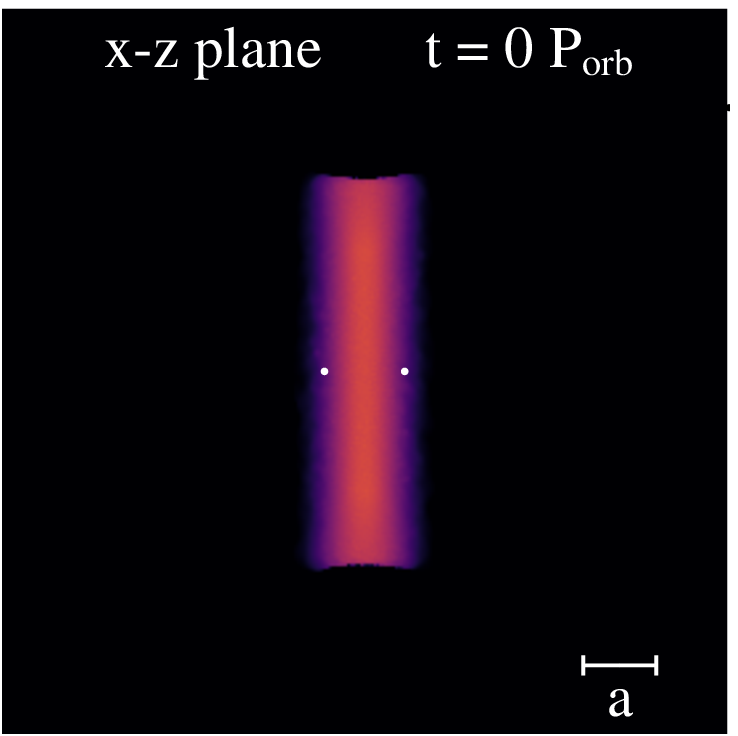}
\includegraphics[width=0.49\columnwidth]{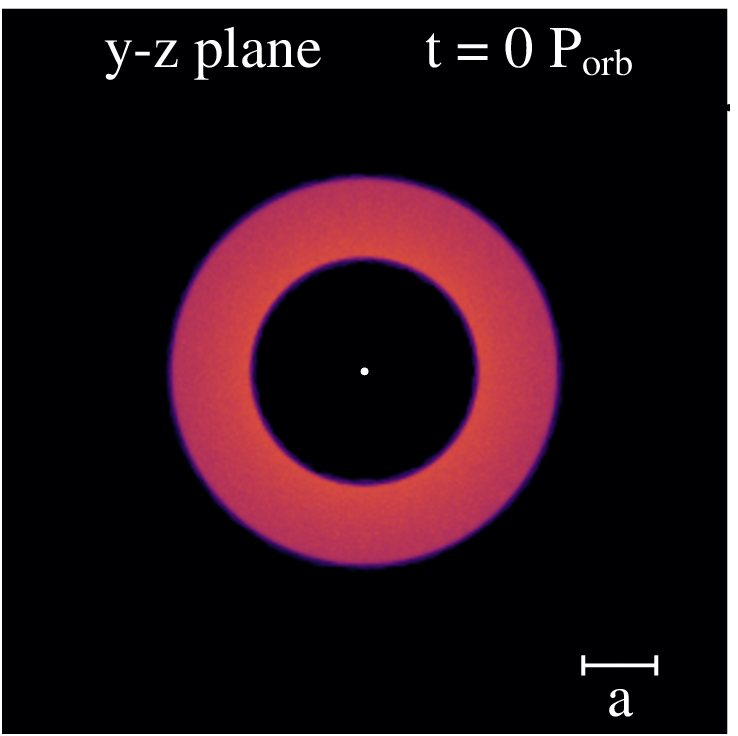} 
\includegraphics[width=0.49\columnwidth]{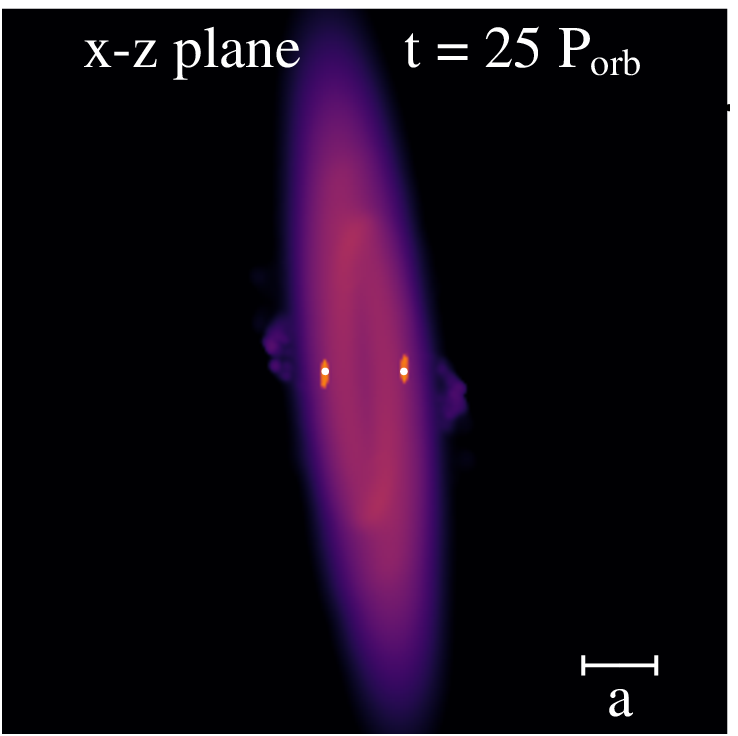} 
\includegraphics[width=0.49\columnwidth]{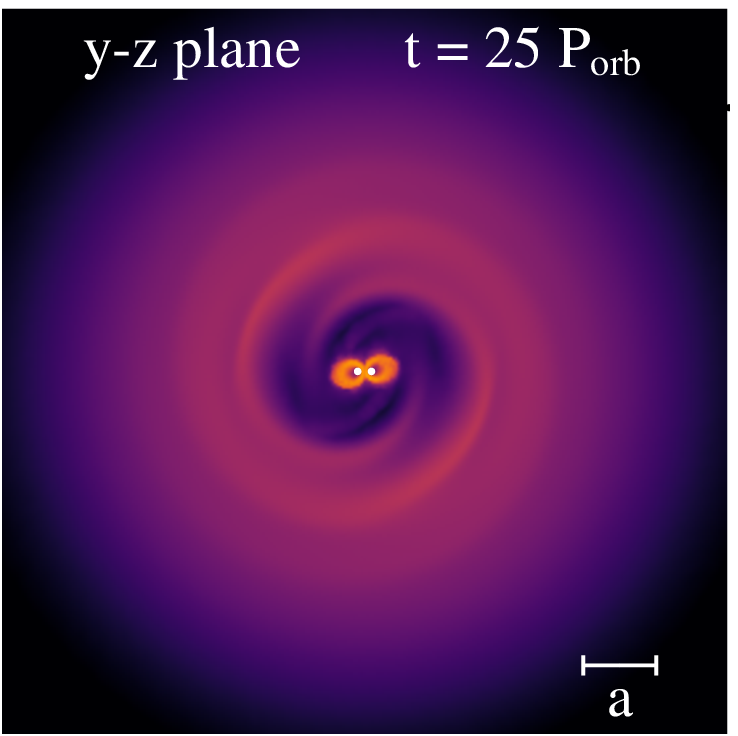} 
\end{center}
\caption{The formation of polar circumstellar discs from an initially low-mass polar circumbinary disc (run8).  The white circles denote the eccentric orbit binary components with an initial binary separation of $a$. The upper panels denote the initial disc setup, while the bottom panels show the disc evolution at $t = 25\, P_{\rm orb}$, where $P_{\rm orb}$ is the binary orbital period. At this time, nearly polar circumstellar discs are forming around each binary component.  The color denotes the gas density using a weighted density interpolation, which gives a mass-weighted line of sight average. The yellow regions are about three orders of magnitude larger than the purple. The left column shows the $x$--$z$ plane, and the right column shows the $y$--$z$ plane.
}
\label{fig::polar_splash}
\end{figure}

\begin{figure} 
\includegraphics[width=\columnwidth]{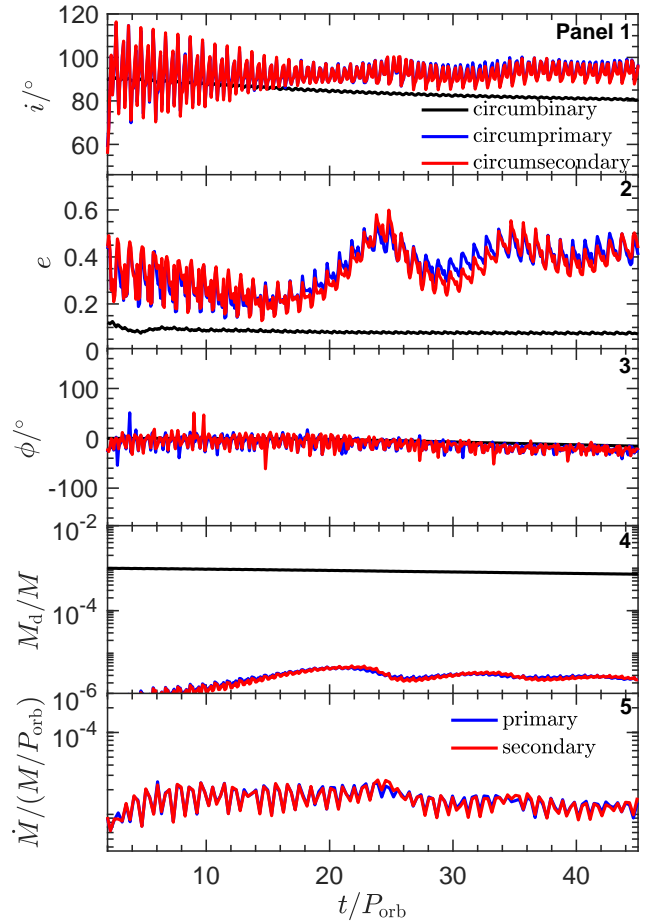}
\centering
\caption{Simulation results for run8  for an initially polar circumbinary disc. The disc parameters are shown for the  circumprimary, circumsecondary, and circumbinary discs as a function of time in units of the binary orbital period, $P_{\rm orb}$. The upper four panels show the  disc tilt $i$ (panel 1), eccentricity $e$ (panel 2), longitude of the ascending node $\phi$ (panel 3), and  disc mass $M_{\rm d}$ (panel 4)  for the three discs. The lower panel shows the mass accretion rate onto the sinks $\dot{M}$ (panel 5).}
\label{fig::i90}
\end{figure}

\begin{figure} 
\includegraphics[width=0.49\columnwidth]{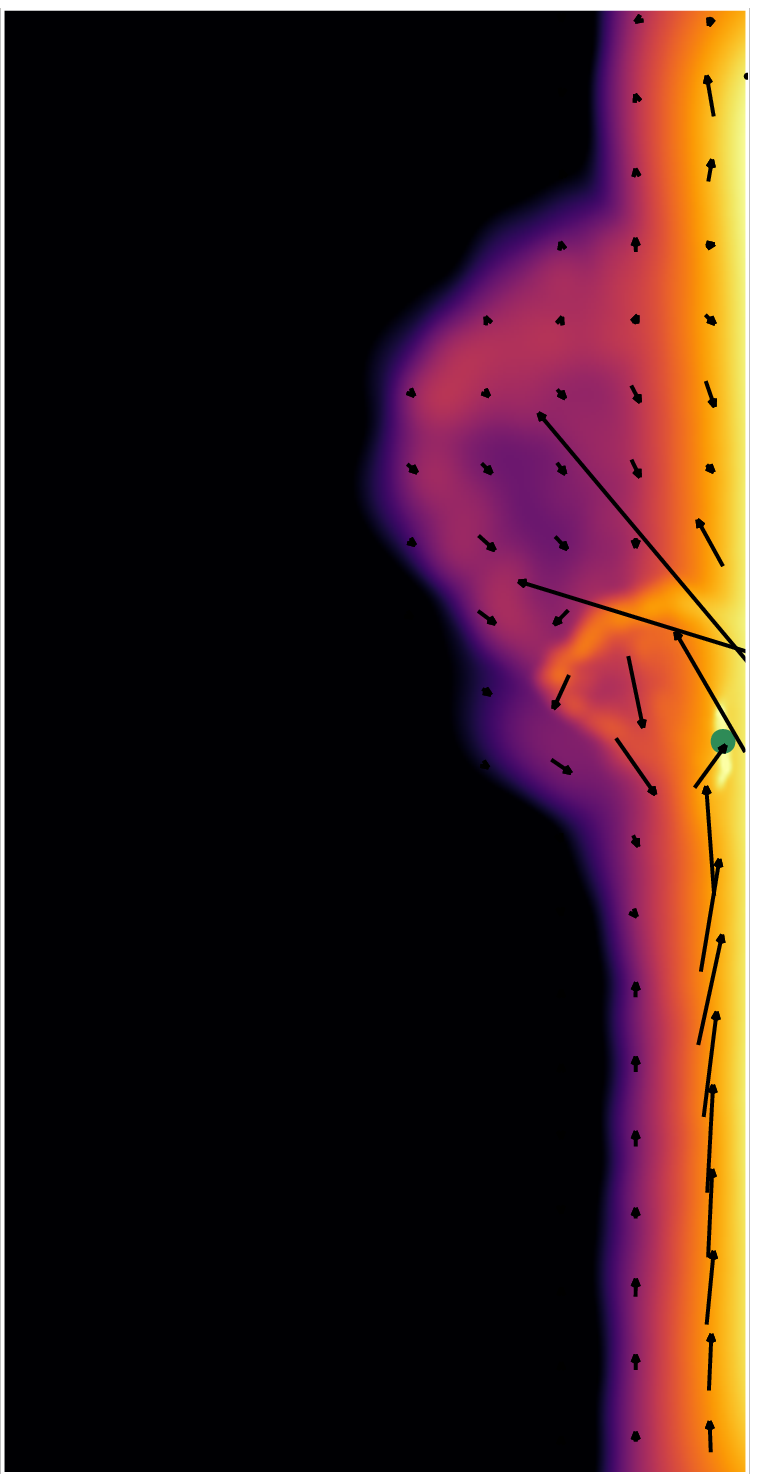}
\includegraphics[width=0.49\columnwidth]{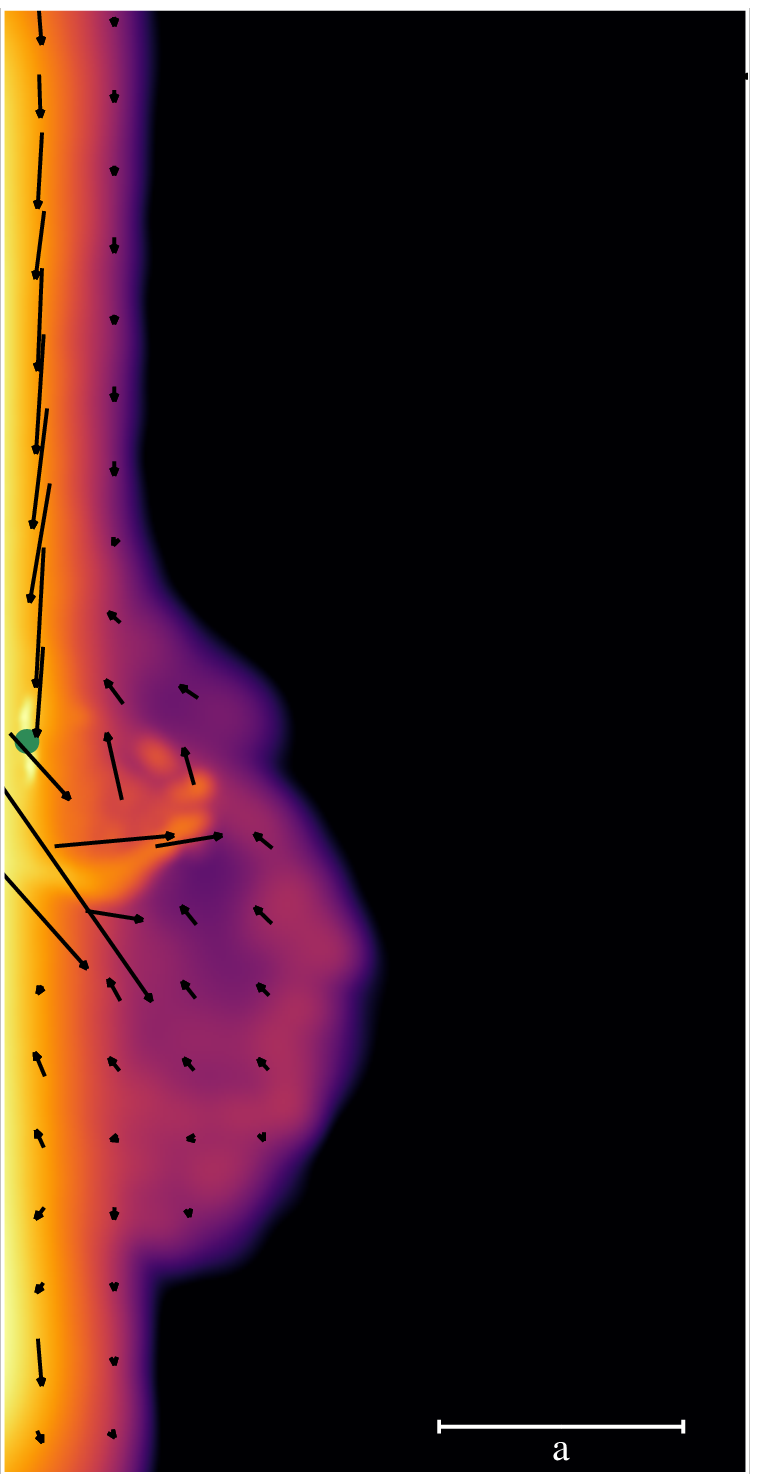}
\centering
\caption{Edge-on view ($x$--$z$ plane) of a polar circumbinary disc (run8) at a time $t = 5\, P_{\rm orb}$. We ignore the main portions of the disc confined within $r < 0.45a_{\rm b}$, where $a_{\rm b}$ is the separation of the binary. The binary components are shown as the green dots. The colours denote the disc surface density, with the orange regions being about three orders of magnitude larger than the purple regions. We overlay the velocity vectors shown by the black arrows. The length of the arrow is proportional to the velocities of the particles. We see two asymmetric lobes of material that are produced by the binary. Several of the velocity vectors are directed away from the plane of the circumbinary disc; however, the material then falls back onto the disc gap.}
\label{fig::vel}
\end{figure}

\subsection{Polar discs}
In this section, we present a hydrodynamical simulation of the flow of material from a polar circumbinary disc onto the binary components (run8). The top row of Fig.~\ref{fig::polar_splash} shows the initial configuration of the polar circumbinary disc around an eccentric binary. The bottom row shows the disc structure at $t = 25\, P_{\rm orb}$. The circumbinary disc remains nearly polar ($\sim 90^\circ$) as shown in the $x$-$z$ plane. Material flows from the polar circumbinary disc and forms nearly polar circumstellar discs around each binary component. The cavity size is smaller in the polar disc  compared to a coplanar disc simulation as expected \citep{Lubow2015, Miranda2015}.

The upper four panels in Fig.~\ref{fig::i90} show the inclination, eccentricity, the longitude of the ascending node, and disc mass for the three discs as a function of time in binary orbital periods. The lower panel shows the mass accretion rate onto the sinks.  The circumstellar discs form at  a time of $\sim 10\, \rm P_{orb}$,  later than in the simulation with a lower level of circumbinary disc misalignment.  The circumbinary disc tilt evolves in time. Since we model a  disc with a non-zero mass, it will align to a generalised polar state  with an inclination that is $<90^\circ$ \cite[e.g.,][]{MartinLubow2019, Chen2019}. In this case, the circumstellar discs form slightly retrograde, with a  tilt just above $90^\circ$. The primary and secondary discs form with an eccentricity of $\sim 0.25$. However, the polar circumstellar discs undergo the KL instability, which forces the disc eccentricity and tilt to oscillate in time.
 Looking at panels 1 and 2, we see that 
as the disc eccentricity increases, the disc tilt also increases, the opposite of the conventional KL case involving prograde orbits. However, this result is consistent with the KL mechanism for retrograde orbits. 
Panel 3 shows the evolution of the longitude of the ascending node in time. Since the circumprimary and circumsecondary discs are nearly polar, they exhibit very little precession  (see equation \ref{omnonKL} and discussion below it). The mass of the polar circumstellar discs oscillates in time (panel 4), likely due to the oscillating disc eccentricity. The polar circumbinary disc has lost $\sim 25$ per cent of its initial mass.

 The KL oscillations from Fig.~\ref{fig::i90} damp in time. However, from our resolution study, the damping is primarily due to the initial number of particles. The accretion timescale for this simulation is $\sim 15\, \rm P_{orb}$, and the KL timescale in this case is $\sim 10\, \rm P_{orb}$. The accretion timescale is longer in the polar simulation than in the $60^\circ$ simulation because the polar circumstellar discs become less eccentric during each KL cycle, accreting less disc material. For a higher resolution, we expect the KL oscillations to be long-lived even for polar circumstellar discs.

On the bottom-left panel in Fig.~\ref{fig::polar_splash}, we see that some material is flung out of the disc plane on both sides of the polar circumbinary disc. This material forms two lobes on both sides of the disc. Figure~\ref{fig::vel} shows the edge-on view of the disc surface density, along with the velocity vectors. The material is being flung outwards but remains bound to the binary. Therefore, the material then falls back into the gap region of the circumbinary disc.  Throughout the simulation, the material is periodically flung out every $0.5\, P_{\rm orb}$ when the binary components pass through the polar circumbinary disc plane.

\begin{figure} 
\includegraphics[width=\columnwidth]{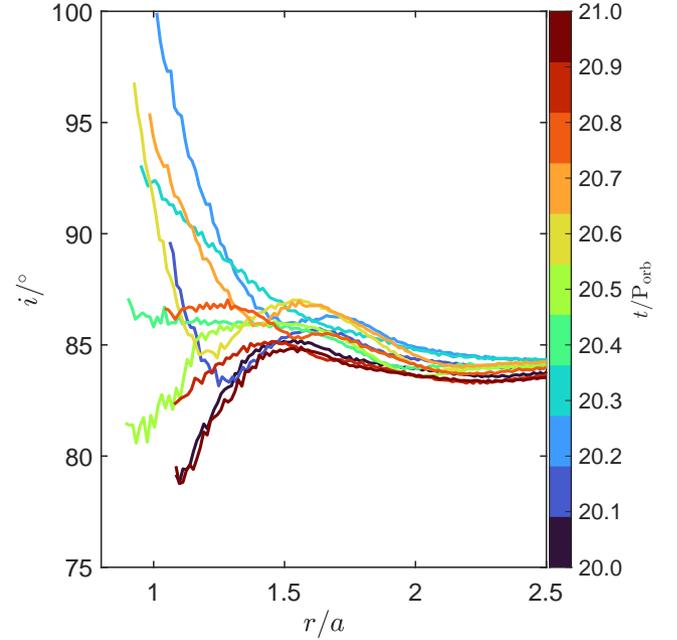}
\centering
\caption{ Circumbinary disc tilt, $i$, as a function of radius, $r$,  for the polar circumbinary disc. The color corresponds to the time in binary orbital periods, $\rm P_{orb}$. }
\label{fig::tilt_radius}
\end{figure}

\begin{figure*} 
\begin{center}
\includegraphics[width=0.49\columnwidth]{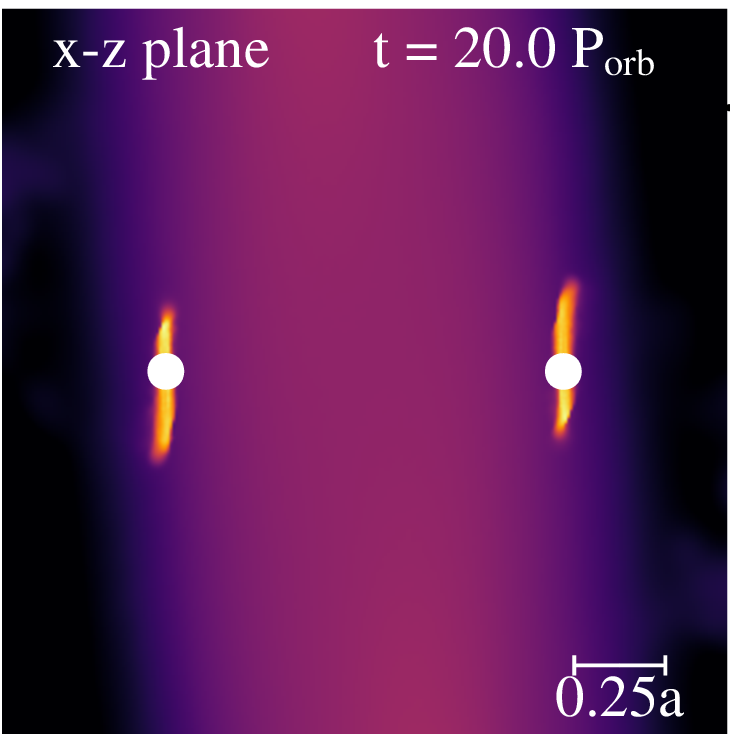}
\includegraphics[width=0.49\columnwidth]{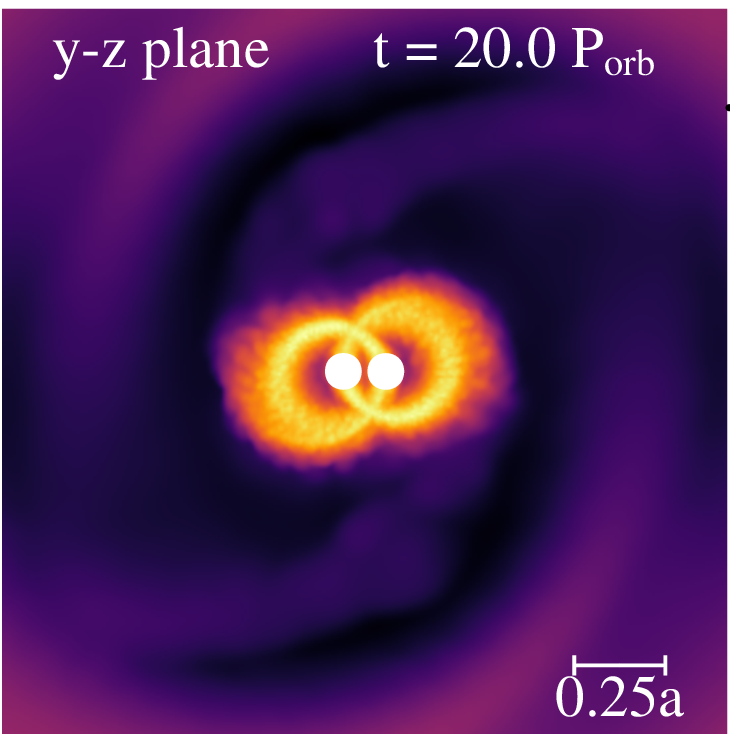}
\includegraphics[width=0.49\columnwidth]{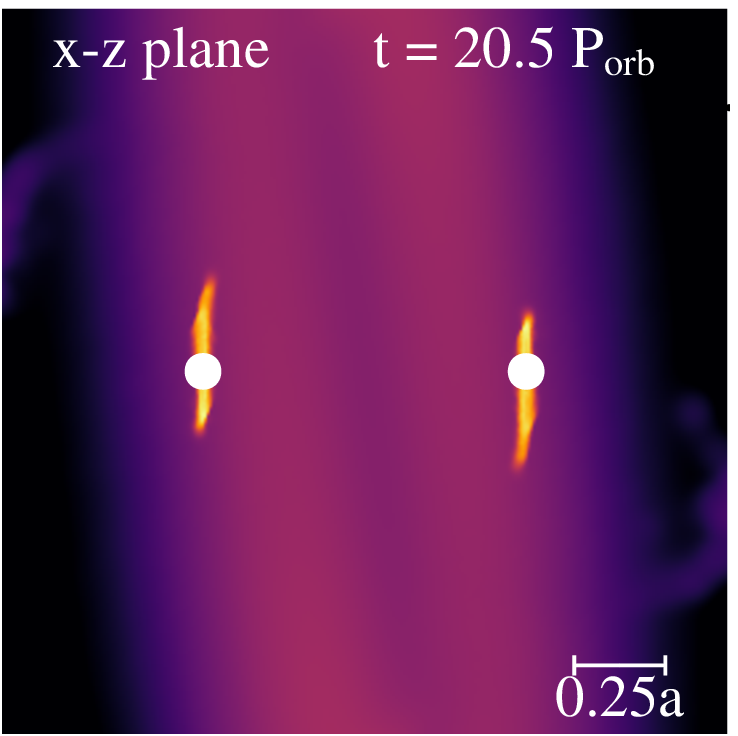}
\includegraphics[width=0.49\columnwidth]{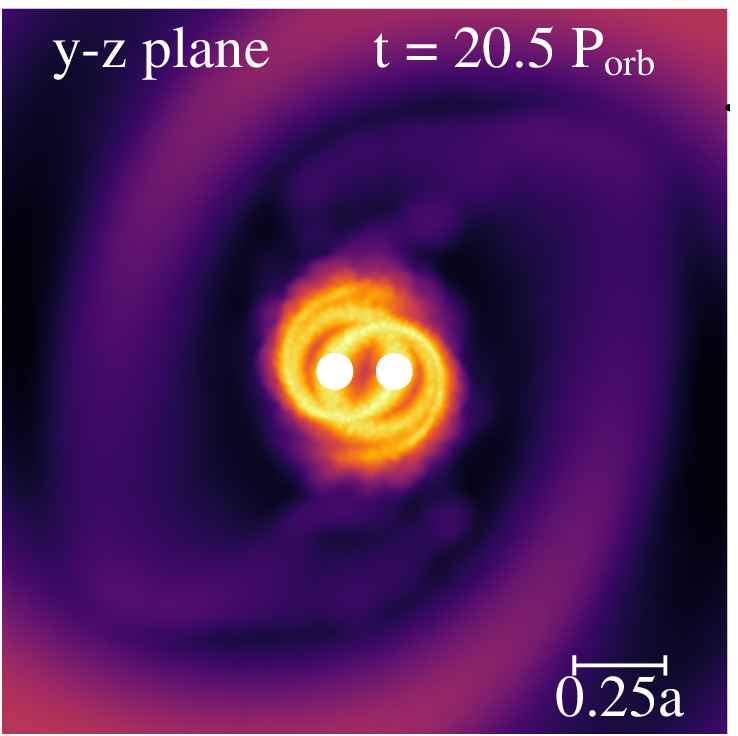}
\includegraphics[width=0.49\columnwidth]{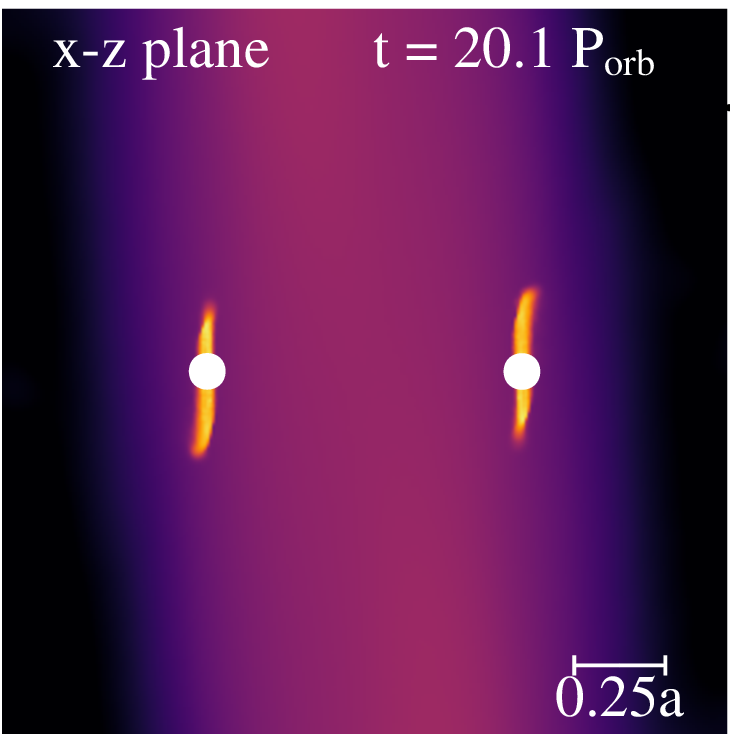}
\includegraphics[width=0.49\columnwidth]{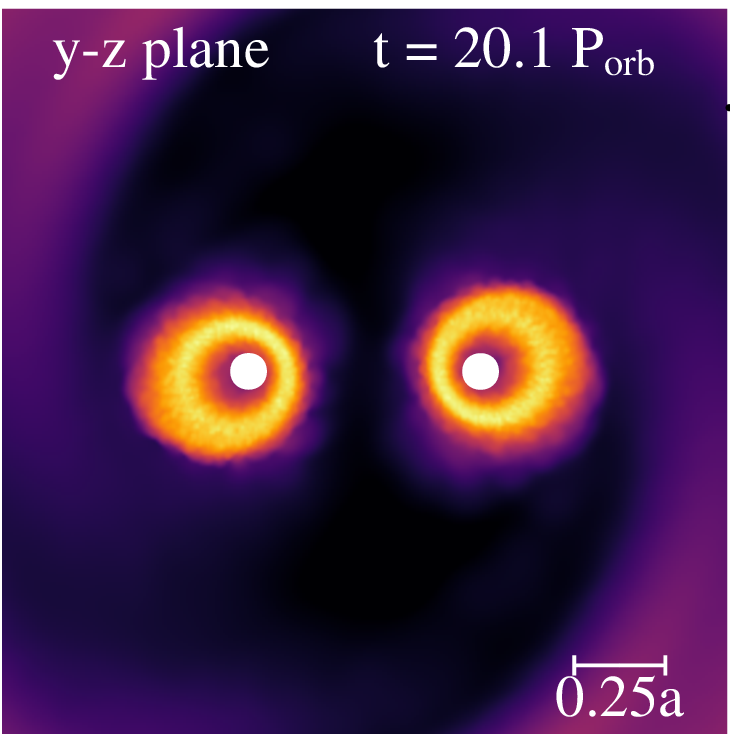}
\includegraphics[width=0.49\columnwidth]{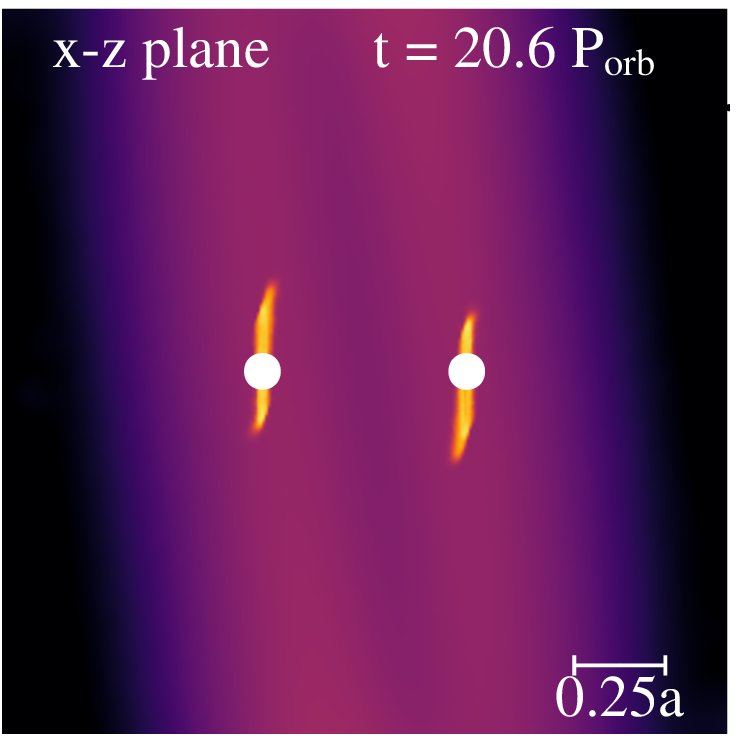}
\includegraphics[width=0.49\columnwidth]{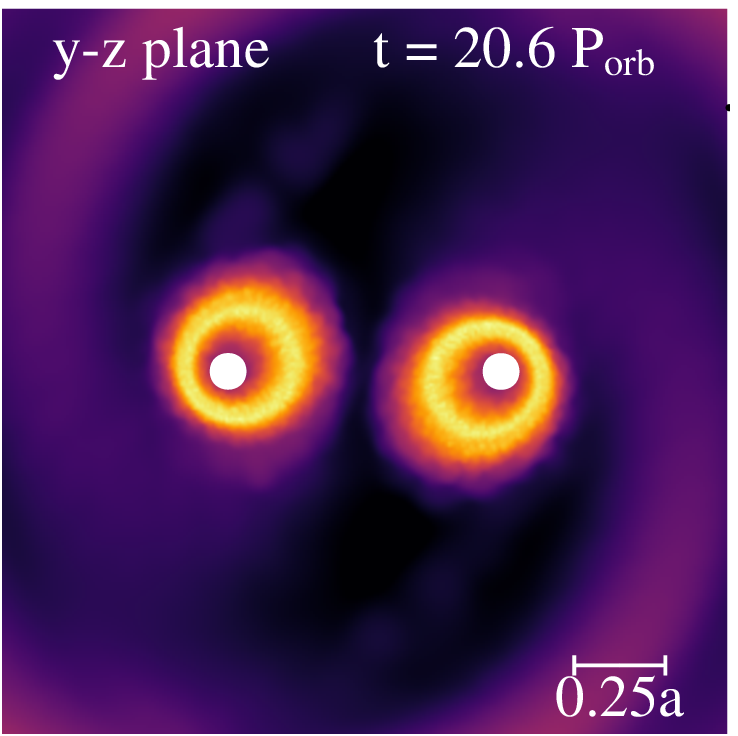}
\includegraphics[width=0.49\columnwidth]{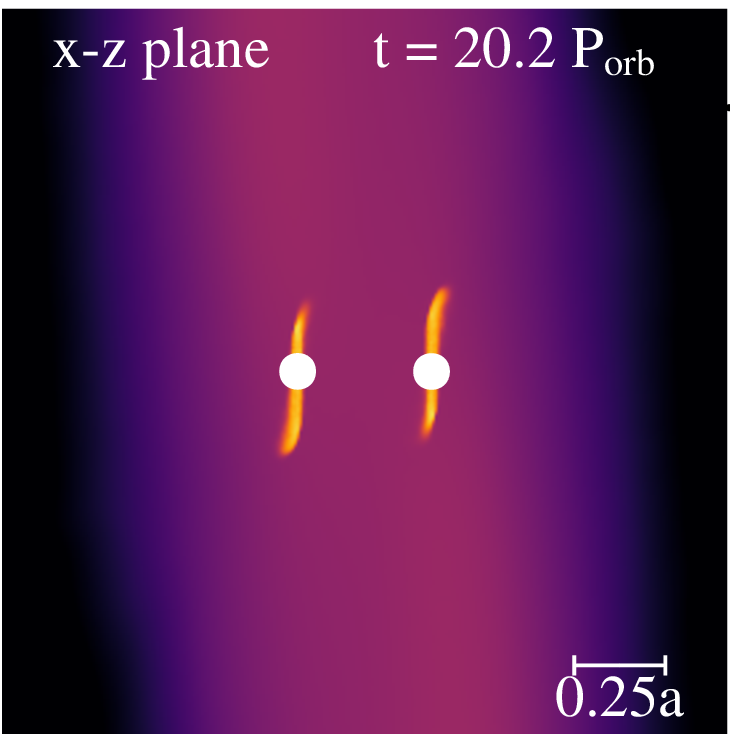}
\includegraphics[width=0.49\columnwidth]{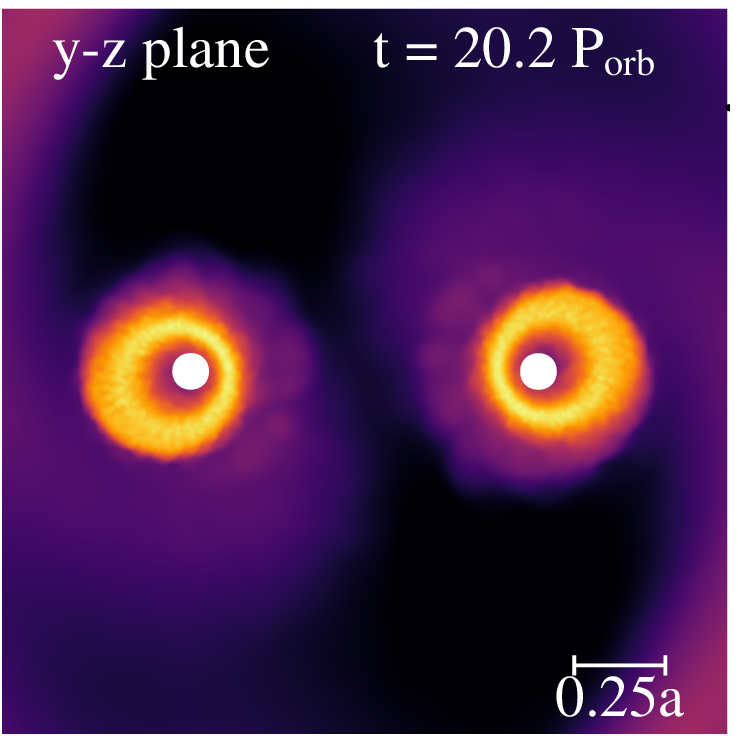}
\includegraphics[width=0.49\columnwidth]{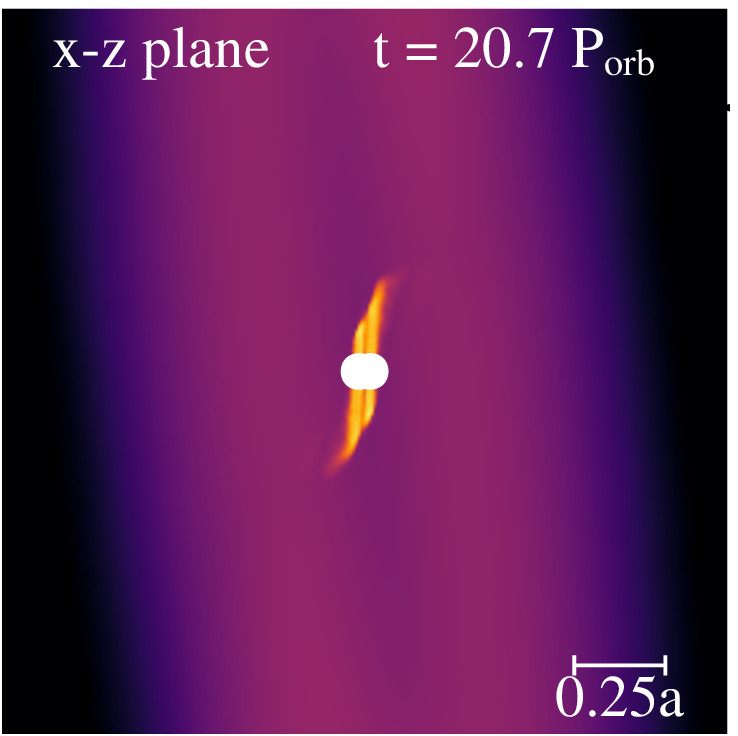}
\includegraphics[width=0.49\columnwidth]{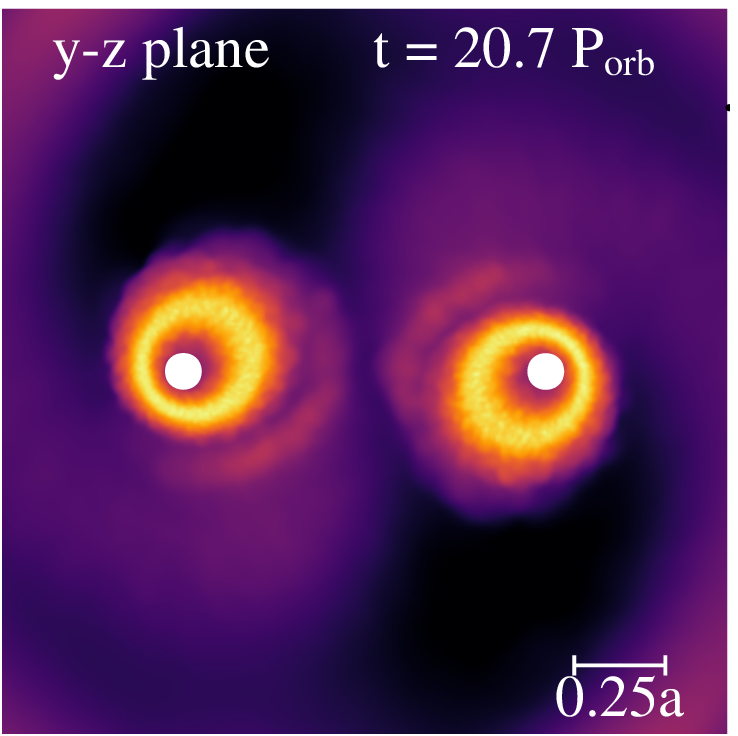}
\includegraphics[width=0.49\columnwidth]{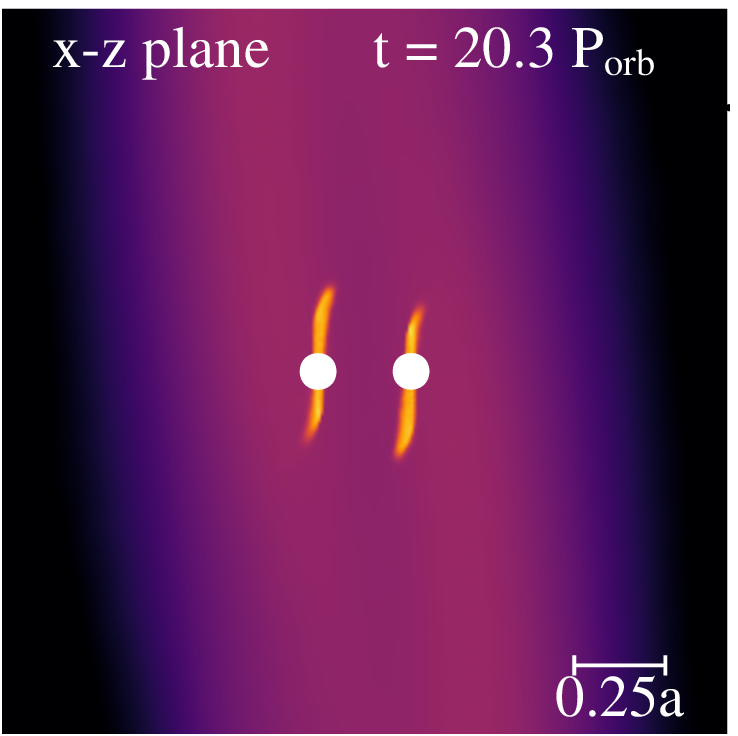}
\includegraphics[width=0.49\columnwidth]{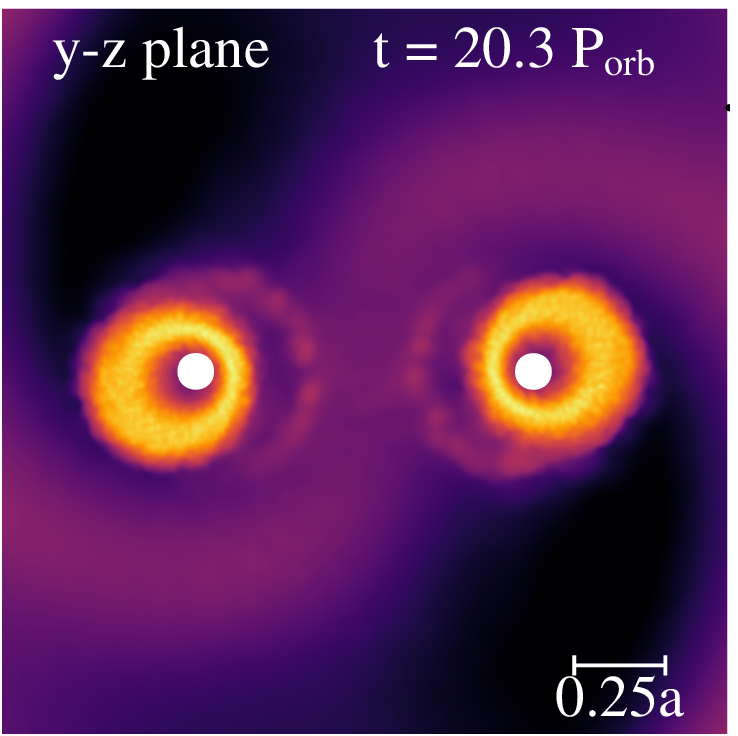}
\includegraphics[width=0.49\columnwidth]{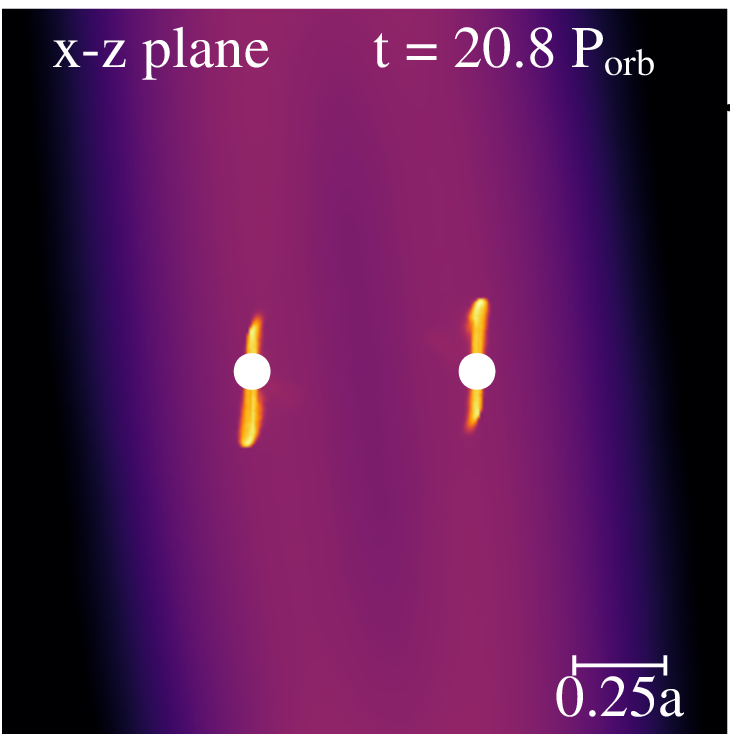}
\includegraphics[width=0.49\columnwidth]{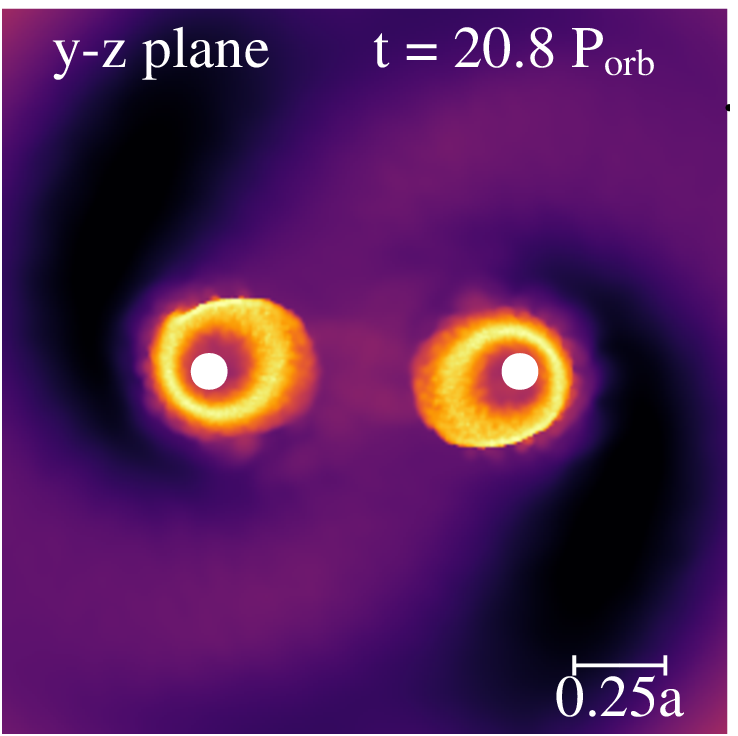}
\includegraphics[width=0.49\columnwidth]{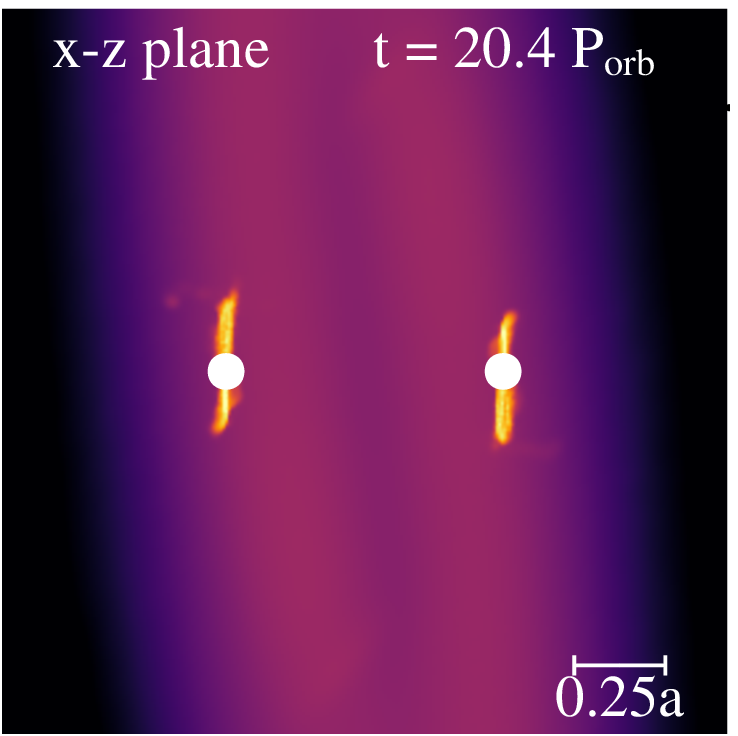}
\includegraphics[width=0.49\columnwidth]{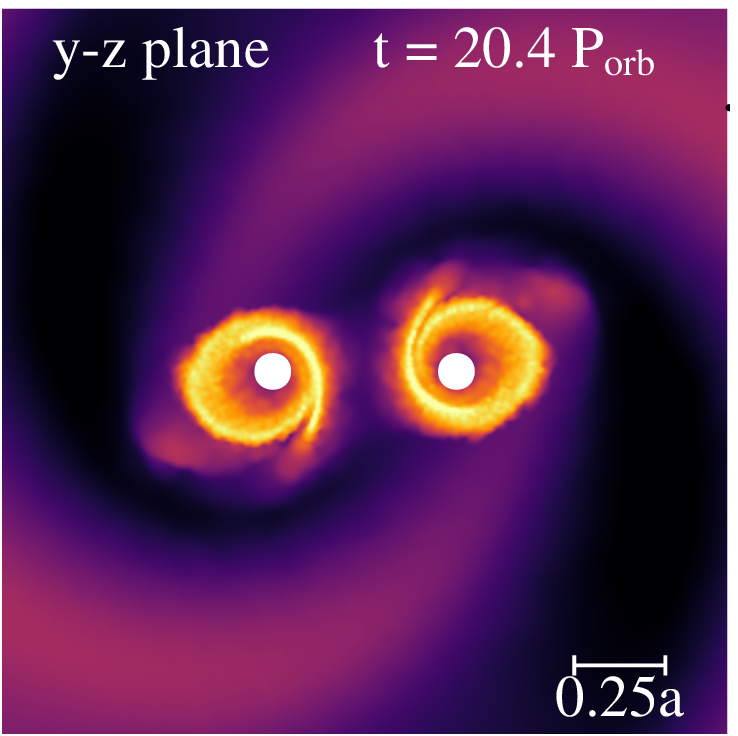}
\includegraphics[width=0.49\columnwidth]{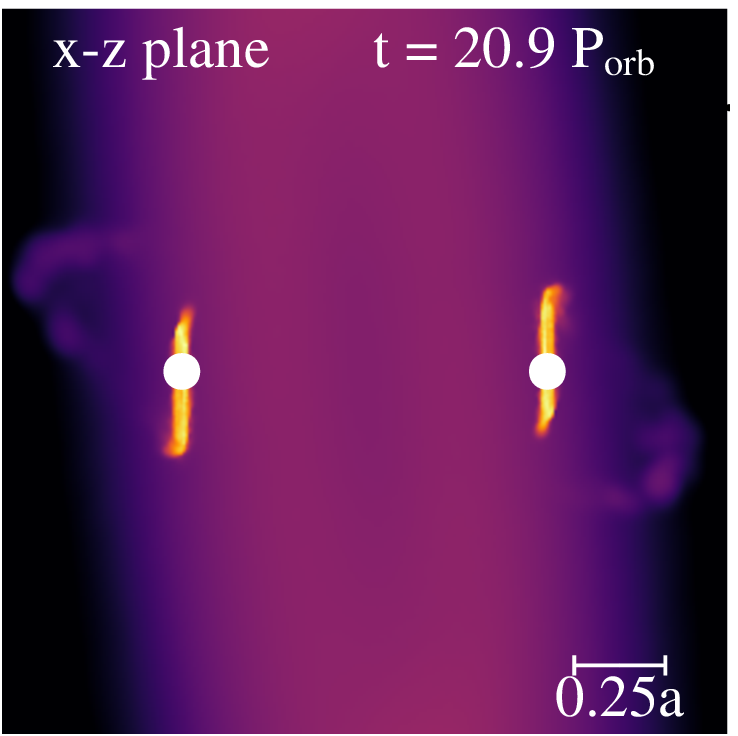}
\includegraphics[width=0.49\columnwidth]{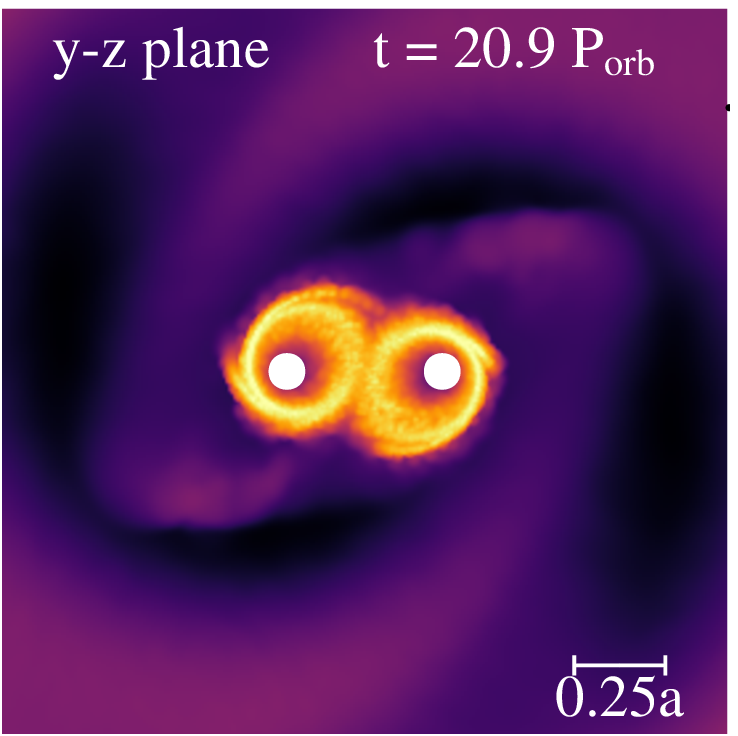}
\end{center}
\caption{Zoomed-in snapshots of the disc surface density showing the flow of material from a polar circumbinary disc onto the nearly polar circumstellar discs. The white circles denote the eccentric orbit binary components with an initial binary separation of $a$.  The color denotes the gas density using a weighted density interpolation, which gives a mass-weighted line of sight average. The yellow regions are about three orders of magnitude larger than the purple. We view the orbit of the binary in the $x$--$z$ and $y$--$z$ planes. The snapshots show a period from $20\, \rm P_{orb}$ to $20.9\, \rm P_{orb}$ in increments of $0.1\, \rm P_{\rm orb}$, where $P_{\rm orb}$ is time in binary orbital periods. }
\label{fig::polar_zoom}
\end{figure*}

\begin{figure} 
\includegraphics[width=\columnwidth]{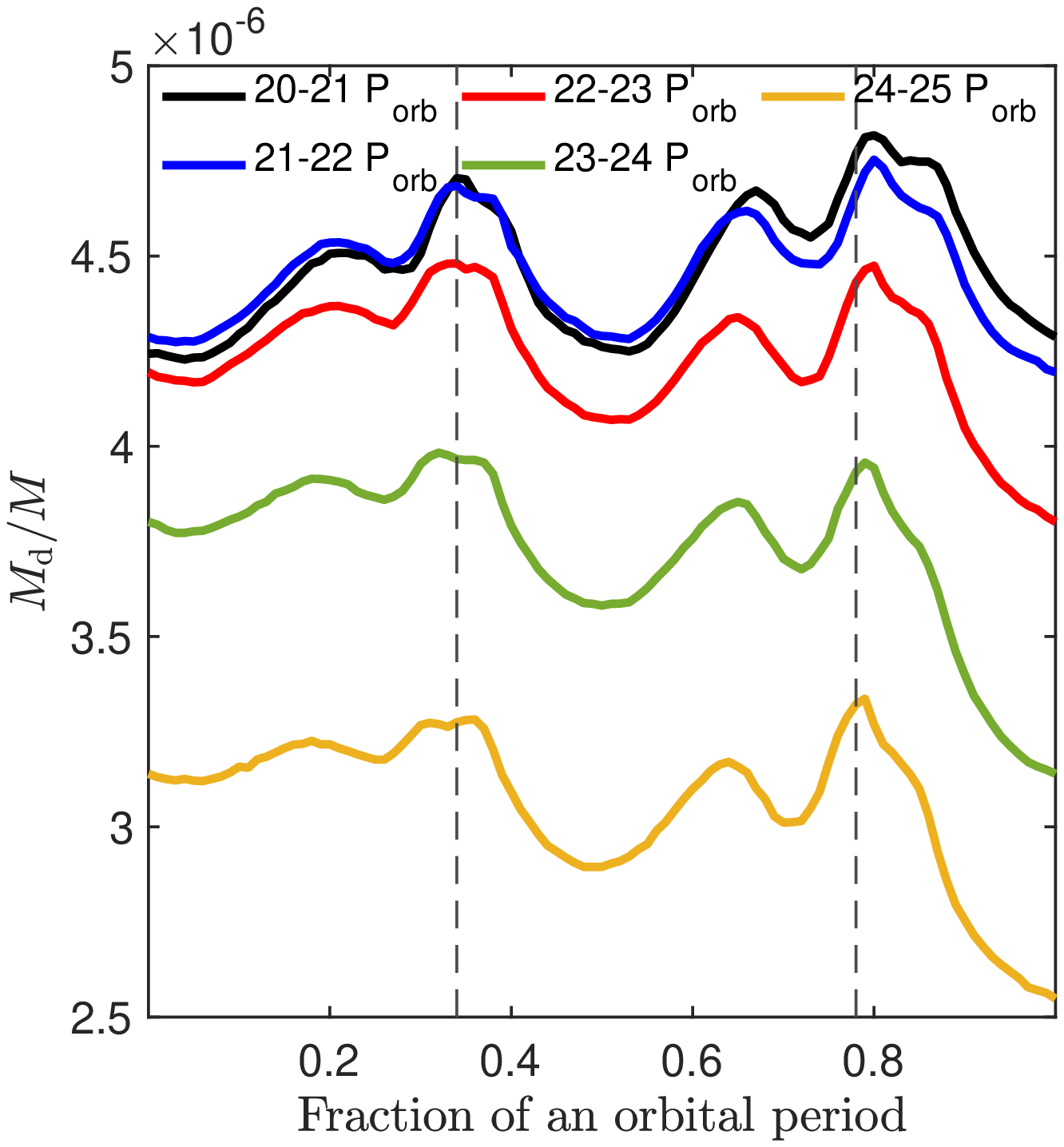}
\centering
\caption{The circumprimary disc mass evolution during one binary orbital period,  $\rm P_{orb}$, at times $20-21\, \rm P_{orb}$ (blue), $21-22\, \rm P_{orb}$ (red), $22-23\, \rm P_{orb}$ (yellow), $23-24\, \rm P_{orb}$ (purple), and $24-25\, \rm P_{orb}$ (green). The mass of the disc decreases every $0.5\, \rm P_{orb}$.   The vertical dashed-lines denote the times when the binary is aligned with the circumbinary disc plane during $20-21\, \rm P_{orb}$. An increased flow of material onto the circumstellar discs occurs when the binary is aligned with the circumbinary disc plane.}
\label{fig::mass_polar_flow}
\end{figure}

We further examine the flow of polar circumbinary material onto the forming circumstellar discs. First, we investigate the tilt of the gaseous streams that accrete onto the circumstellar discs as a function of time. Figure~\ref{fig::tilt_radius} shows the  circumbinary disc tilt as a function of disc radius. The inner edge of the disc lies roughly at $1.6a$. The curves that are shown at radii $<1.6a$ map the tilt of the streams. We show the disc tilt for a full binary orbital period from $20\, \rm P_{orb}$ to $21\, \rm P_{orb}$ in increments of $0.1\, \rm P_{orb}$. At every $0.5\, \rm P_{orb}$, the tilt of the streams are low at $\sim 80^\circ$. When the binary orbital period is not at half increments, the tilt of the streams increases beyond $90^\circ$. For example, at times $20.2-20.3\, \rm P_{orb}$ and $20.6-20.7\, \rm P_{orb}$, the streams are highly tilted. Recall that the forming circumstellar discs initially form at a high disc tilt, $> 90^\circ$. Therefore, whenever the gaseous streams are highly tilted, there is an increased accretion of material onto the circumstellar discs from the circumbinary disc. When the streams are less inclined, every $0.5\, \rm P_{orb}$, there will be less material accreted onto the polar circumstellar discs. This  phenomenon is also consistent with Fig.~\ref{fig::vel}, where material is flung out of the plane of the circumbinary disc every $0.5\, \rm P_{orb}$. We test this by further visualizing the inflow of material. Figure~\ref{fig::polar_zoom} shows snapshots of zoomed-in views in the $x$--$z$ and $y$--$z$ planes of the disc surface density, showing the gaseous streams accreting onto the nearly polar circumstellar discs. The snapshots show the flow of material over $20\, \rm P_{orb}$ to $20.9\, \rm P_{orb}$ in increments of $0.1\, \rm P_{orb}$. Higher density streams occur at times $20.3\, \rm P_{orb}$ and $20.7\, \rm P_{orb}$. The flow of material decreases every $0.5\, \rm P_{orb}$ during the orbit. At these times, the steams are less dense, leading to less material accreting onto the circumstellar discs.

 We relate the flow of material from Fig.~\ref{fig::polar_zoom} to the mass of the circumstellar discs. Figure~\ref{fig::mass_polar_flow} shows the mass of the circumprimary disc from $20\, \rm P_{orb}$ to $25\, \rm P_{orb}$ folded on top of one another for each orbital period.   The vertical dashed-lines denote the times when the binary is aligned with the circumbinary disc plane, which is assumed when the stars are both aligned with $x$--$z$ plane. Each time the binary aligns to the plane of the disc, the  masses of the circumstellar discs increase. The mass of the disc decreases every $0.5\, \rm P_{orb}$.  This behaviour repeats every orbital period. Overall, the disc mass deceases in time due to the KL mechanism. 

\section{Summary}

%\RGM{Main conclusions seem to be:
%1. A CBD with inclination $\gtrsim 40$ forms circumstellar discs that are KL unstable.
%2. Even if the KL oscillations do damp out, the disc remains highly eccentric (e=0.6) with the source of high inclination material.
%3. A polar CBD forms almost polar CSDs that undergo KL.
%}
\label{sec::summary}

In this work, we investigated the flow of material from a circumbinary disc that results in the formation circumstellar discs around each binary component. We simulated an initially highly misaligned and polar circumbinary disc using three-dimensional SPH. We considered cases of low initial binary eccentricity (typically $e_{\rm b}=0.1$)
and binary mass ratio of unity. We also simulated cases of test particles around the primary star  and cases of circumprimary discs only (i.e., no circumbinary or circumsecondary discs) for comparison. 

In order to carry out these simulations in a reasonable amount of time, we made some compromises on our choice of parameters. In particular, we introduced a higher viscosity parameter for the circumbinary disc than is likely to occur and a lower temperature of the gas in the gap region. These choices were made to improve the resolution of the simulations. Even with these parameters, the resolution is still playing a role in our results (see Fig.~\ref{fig::i60}).   While we have chosen the disc parameters ($\alpha$ and $H/R$) in our simulations to maximise the accretion rate on to the binary components and therefore the simulation resolution, we expect the general behaviour to persist for more realistic parameters applicable to protoplanetary discs.    The mass of the circumstellar disc scales with the infall accretion rate. If the resolution of the circumstellar disc is too poor, then the disc artificially accretes rapidly due to the artificially enhanced effects of viscosity at low density in the SPH code.  

We first examined the behavior of initially highly inclined circumstellar discs
that are not supplied with material from a circumbinary disc.
 A polar test particle in orbit around a primary star reaches an eccentricity of nearly unity during the first KL cycle, forcing the particle to become unbound or hit the central star.   Similarly, initially highly inclined circumstellar discs around individual binary components can experience  very strong KL oscillations. For an equal mass binary containing only a single circumstellar disc at high inclination between $70^{\circ}$ and $100^{\circ}$, the disc undergoes  only a single KL oscillation
before losing nearly all its mass  for our given sink size.  Some of the disc mass is transferred to the companion star to form a low inclination disc that does not undergo KL oscillations. These results suggests that such high inclinations
of discs are short-lived due to  enhanced dissipation from shocks that leads to tilt evolution on short timescales. In contrast, discs  that are  highly inclined but are not subject to KL  oscillations  would undergo much slower evolution. In particular, a polar disc would  not precess (see e.g., equation (\ref{omnonKL})) and therefore not warp. The disc would then not be subject to torques that act to change its inclination.  

In this work, and from \cite{Smallwood2021}, we showed that the continuous accretion of material from the circumbinary disc allows   the effects of KL  oscillations  on  circumstellar discs to be much longer-lived. In this process, the circumbinary material is continuously delivered with a high inclination to the lower inclination circumstellar discs. We found that the simulation resolution is important for modeling the longevity of the KL oscillations. We find longer lived KL oscillations that show signs of mild weakening in time, possibly due to the resolution (e.g., Figure \ref{fig::i60}). The balance between the accretion timescale and the KL timescale determines whether the oscillations are sustained or damp in time. If the circumstellar disc material were to accrete on a much shorter timescale than the KL oscillation period, we would not expect the KL oscillations to operate. We found that with increasing resolution, the accretion timescale becomes comparable to the KL timescale, favoring sustained KL oscillations. 

%We find that {\bf in some cases} the longer-lived oscillations evolve to a stable, highly eccentric disc for  a circumbinary disc with initial tilt $\gtrsim 60^\circ$.{\bf We caution that the stability is found over the limited time of about 20 binary orbit periods.}
% For highly misaligned circumbinary discs around unequal-mass binaries, the forming circumstellar discs also become KL unstable and evolve to a stable eccentric disc. 

%Next, we examined the dynamics of polar discs and inclined test particles in binary star systems. A polar test particle around a primary star reaches an eccentricity of $e = 1$ during the first KL cycle, forcing the particle to become unbound or hit the central star. A hydrodynamical simulation of a system containing only a polar circumprimary disc shows that this
%disc loses its mass after the first KL cycle as noted above.  This situation changes in the presence of a misaligned circumbinary disc.
%An initially polar circumbinary disc results in the formation of nearly polar circumstellar discs around each binary component that are long-lived and undergo KL oscillations. The KL oscillations in the simulations are long-lived provided that the resolution is high enough for the accretion timescale to be comparable to the KL timescale. Sustained KL oscillations will cause the polar circumstellar discs to maintain a non-zero disc eccentricity as long as the KL mechanism operates. 

Planet formation is thought to still occur in non-zero eccentricity discs \citep{Silsbee2021}. In the case of {\it S}-type planets (planets orbiting one of the stellar companions in a binary), gravitational perturbations from an  eccentric orbit stellar companion and an eccentric disc increase planetesimal eccentricities, leading to collisional fragmentation, rather than growth, of planetesimals. However, \cite{Rafikov2015} analyzed the planetesimal motion in eccentric protoplanetary discs when the planetesimals were affected by gas drag and disc gravity. They found that the planetesimals could withstand collisional fragmentation and erosion, thereby providing a pathway to forming planetary cores by coagulation in a binary. It is not clear how those results carry over to the case of highly eccentric discs undergoing KL oscillations. However, the formation of nearly polar circumstellar discs from this work may give rise to the formation of nearly polar planets that become Kozai-unstable.  Planet formation in a polar circumstellar disc requires the disc to last for a sufficiently long time. We speculate that this is possible provided that the disc is continuously accreting material in a polar configuration.   

Observations of misaligned planetary systems show a preference for nearly polar orbits with true obliquities $\psi$ in the range $\psi =80^\circ - 125^\circ$ \citep{Albrecht2021,Dawson2021}. For example, two observed  ultra-short-period hot Jupiters in polar orbits around an A-type star are Kelt-9b \citep{Ahlers2020b} and MASCARA-4b \citep{Ahlers2020a}.  The majority of planets studied by \cite{Albrecht2021} were hot Jupiters, since the measurements for these types of planets are more precise. However, a few warm-Neptunes with polar orbits were observed, including  HAT-P-11b \citep{Sanchis-Ojedda2011}, GJ 436b \citep{Bourrier2018,Bourrier2022},  HD 3167c \citep{Dalal2019,Bourrier2021}, and WASP-107b \citep{Dai2017,Rubenzahl2021}. A more recent warm Neptune, GJ 3470b, is also observed to be on a polar orbit \citep{Stefansson2022}.

%An initially polar circumbinary disc forms nearly polar circumstellar discs around each binary component. The circumstellar discs undergo KL oscillations that are more long-lived compared to simulations without accretion, but ultimately they evolves to a stable eccentric polar circumstellar disc. Therefore, the continuous accretion of polar circumbinary material on to the polar circumstellar discs causes the KL mechanism to be sustained over many KL oscillation periods. The KL unstable polar circumstellar discs may give rise to the formation of polar S-type planets. \RGM{I'm not sure I'd say "many KL periods". I'd rather emphasize the highly eccentric polar disc. Planet formation would have to take place in an eccentric disc. Perhaps add some citations/discussion for this. e.g. Rafikov? }

\section*{Acknowledgements}
 We thank the anonymous reviewer for helpful suggestions that positively impacted the work. We thank Daniel Price for providing the {\sc phantom} code for SPH simulations and acknowledge the use of SPLASH \citep{Price2007} for the rendering of the figures. Computer support was provided by UNLV's National Supercomputing Center. We acknowledge support from NASA XRP grants 80NSSC19K0443 and 80NSSC21K0395. This research was supported in part by the National Science Foundation under Grant No. NSF PHY-1748958. SHL thanks the Simons Foundation for support during a visit to the Flatiron Institute.

%%%%%%%%%%%%%%%%%%%%%%%%%%%%%%%%%%%%%%%%%%%%%%%%%%
\section*{Data Availability}
The data supporting the plots within this article are available on reasonable request to the corresponding author. A public version of the {\sc phantom}, {\sc splash}, and {\sc mercury} codes are available at \url{https://github.com/danieljprice/phantom}, \url{http://users.monash.edu.au/~dprice/splash/download.html}, and \url{https://github.com/4xxi/mercury}, respectively.



\bibliographystyle{mnras}
\bibliography{ref} % if your bibtex file is called example.bib

%\begin{thebibliography}{99}
%\end{thebibliography}

%%%%%%%%%%%%%%%%%%%%%%%%%%%%%%%%%%%%%%%%%%%%%%%%%%

%%%%%%%%%%%%%%%%% APPENDICES %%%%%%%%%%%%%%%%%%%%%

%\appendix


%%%%%%%%%%%%%%%%%%%%%%%%%%%%%%%%%%%%%%%%%%%%%%%%%%

% Don't change these lines
\bsp	% typesetting comment
\label{lastpage}
\end{document}